\documentclass[twocolumn,superscriptaddress]{revtex4-1}
\usepackage{graphicx}
\usepackage{epstopdf}
\usepackage{amsmath}
\usepackage{booktabs}
\usepackage{color}
\usepackage{multirow}
\begin{document}

\title{Crystal nucleation as the ordering of multiple order parameters}

\author{John Russo}
\email{russoj@iis.u-tokyo.ac.jp} 
\affiliation{ {Institute of Industrial Science, University of Tokyo, Meguro-ku, Tokyo 153-8505, Japan} }
\affiliation{ {School of Mathematics, University of Bristol, Bristol BS8 1TW, United Kingdom} }
\author{Hajime Tanaka}
\email{tanaka@iis.u-tokyo.ac.jp}
\affiliation{ {Institute of Industrial Science, University of Tokyo, Meguro-ku, Tokyo 153-8505, Japan} }

\begin{abstract}
Nucleation is an activated process in which the system has to overcome a free energy barrier in order for a first-order phase transition between the metastable and the stable phases to take place. In the liquid-to-solid transition the process occurs between phases of different symmetry, and it is thus inherently a multi-dimensional process, in which all symmetries are broken at the transition. In this Focus Article, we consider some recent studies which highlight the multi-dimensional nature of the nucleation process. Even for a single-component system, the formation of solid crystals from the metastable melt involves fluctuations of two (or more) order parameters, often associated with the decoupling of positional and orientational symmetry breaking. In other words, we need at least two order parameters to describe the free-energy of a system including its liquid and crystalline states. This decoupling occurs naturally for asymmetric particles or directional interactions, focusing here on the case of water, 
but 
we will show that it also affects spherically symmetric interacting particles, such as the hard-sphere system. We will show how the treatment of nucleation as a multi-dimensional process has shed new light on the process of polymorph selection, on the effect of external fields on the nucleation process, and on glass-forming ability.
\end{abstract}

\maketitle

\section{Introduction}

During a first-order phase transition, the system has to overcome a free energy barrier in order to transform to
the thermodynamically stable phase. When this process is driven only by thermal fluctuations, it is called
\emph{homogeneous nucleation}. More often, nucleation is facilitated by external surfaces and the corresponding
process is known as \emph{heterogeneous nucleation}.
Nucleation is an important process for its many practical applications, as it is ubiquitous
in various types of technologically important materials such as metals, semiconductors, pharmaceuticals, and foods.
Nucleation has been thoroughly studied (for recent reviews see for example Refs.~\cite{kelton2010nucleation,sosso2016crystal}), but it is still not entirely understood,
due to the experimental difficulty in accessing the small size
of the crystalline nuclei involved in the transition (which is of the order of 10-1000 molecules in usual conditions).
The difficulty in observing the nucleation process at the microscopic level has profound repercussions not 
only on our theoretical understanding, but also on the practical task of estimating nucleation rates,
where considerable disagreements between experimental measurements and numerical predictions are often
reported even for the simplest systems.

One notable example is the condensation of Argon, a noble gas,
where reports of theoretical and experimentally measured nucleation rates can differ even by 26 orders of magnitude~\cite{iland2007argon,kalikmanov2008argon,sinha2010argon}.

A molecular system whose crystallization represents the most important physical transformation on Earth is water.
Understanding nucleation of water is of fundamental importance in many industrial and natural processes.
For example, the nucleation of ice nuclei in atmospheric clouds is an important factor in establishing
Earth's radiation budget. Our lack of understanding of the freezing mechanisms of water in clouds has
a negative impact on many climate models: ice crystals control the amount of light reflected by clouds,
which is one of the largest unknown parameters that affect atmospheric warming~\cite{stocker2014climate}

Differently from atomic and molecular systems, colloidal systems are characterized by lengthscales
(of the order of $100 \sim 1000$ $\mu$m) and timescales (of the order of seconds) that can be
accessed at a single particle level with microscopy techniques. Moreover, the interactions between colloids
are mediated by fast degrees of freedom (like the solvent); these effective interactions~\cite{likos}
can be controlled to a high degree by changing a few external parameters, like the quality and concentration of the
dissolved salts. These considerations give to colloidal suspensions the quality of ideal systems, where
our theoretical understanding of crystallization can be benchmarked against accurate experimental results.
For a recent review of nucleation in colloidal systems see Ref.~\cite{russo2016nonclassical}.
It is surprising to observe that even for the simplest colloidal system,
i.e. colloidal hard spheres~\cite{gasser,lowen2007critical,assoud2009crystal,leocmach2012roles,jade_royall}, discrepancies between the observed and predicted nucleation
rates can be as big as ten orders of magnitude~\cite{auer2001prediction}.
While numerical simulations have found a rapid increase of the nucleation
rate with increasing colloid volume fraction $\phi$, growing by more than 15 orders of magnitude from
$\phi=0.52$ to $\phi=0.56$~\cite{auer2001prediction,zaccarelli,filion,kawasaki,kawasaki2010correction,kawasaki2010structural,pusey2009hard,filion2,schilling_jpcm,valeriani2012compact,palberg2014crystallization},
experiments on both polymethyl methacrylate and polystyrene microgel systems, instead report nucleation rates
that are much less sensitive to volume fraction~\cite{schatzel1993density,harland1997crystallization,sinn2001solidification,iacopini,franke2011heterogeneous}.

The simplest and most general understanding of nucleation is embodied in Classical Nucleation Theory (CNT)~\cite{becker1935kinetische,zeldovich1943theory,kelton2010nucleation}.
Focusing here on the liquid-to-solid transition, the process starts from small crystalline nuclei that spontaneously form
in the supercooled liquid from thermal fluctuations. At any temperature below the melting temperature, there is a
characteristic size at which these nuclei are equally likely to dissolve or to grow until the liquid-to-solid
transition is complete. This size is called \emph{critical nucleus size}. Classical Nucleation Theory
rests on the assumption that the critical nucleus is amenable to a thermodynamic description, and that
the process is effectively one-dimensional, meaning that it can be described by just one reaction coordinate,
with all order parameters involved in the transition proceeding simultaneously.

The idea of describing crystal nucleation as a one-dimensional process is borrowed from first-order phase
transitions between phases with the same symmetry, most notably the gas-liquid nucleation, or its reverse
process, cavitation. But the liquid-to-solid phase transition is different, as both the translational and
orientational symmetries of the liquid are broken with crystallization.
This is seen by expanding the density of the crystal in its Fourier components~\cite{oxtoby}:
$$
\rho(\mathbf{r})=\rho_l(1+\eta)+\rho_l\sum_n\mu_n\exp(i\mathbf{k}_n \cdot \mathbf{r}),
$$
where $\rho_l$ is the density of the liquid phase, $\eta=(\rho_s-\rho_l)/\rho_l$ is the fractional density
change at freezing, and $\mu_n$ are the Fourier components associated with the crystalline periodic structure
($\mu_n=0$ for the liquid phase). The liquid-to-solid transition is thus associated with an infinite
number of order parameters ($\eta$, $\mu_n$). By assuming the crystal to be harmonic at melting, all high
order order parameters can be related to $\mu_1$. In this effective case, the transition is thus represented
by two order parameters, $\eta$ and $\mu_1$, which are associated with translational and orientational order respectively. 
We note, however, that the latter still includes translational order due to the constraint on the wave vector, $\mathbf{k}=\mathbf{k}_n$, 
which is a feature of the density functional theory of crystallization.

The fact that nucleation is inherently multi-dimensional is obvious in all those systems where the breaking
of positional and orientational order occurs at different temperatures, the most famous example being
the freezing of two dimensional hard disks. Here, between the liquid and solid phases, an intermediate phase
appears, called the hexatic phase, characterized by quasi-long range orientational order, but short-range positional
order. In one of the most important cases of algorithmic development in recent years~\cite{bernard2011two}, it was shown that the
transition between the liquid and hexatic phase is first-order, while the hexatic-to-solid is a second order transition.

But a decoupling of translational and orientational order can occur even when both symmetries break down simultaneously as in the case of crystallization in three dimensions. 
Note that translational ordering automatically accompanies orientational ordering. In particular, in the next sections we show examples
where two-step ordering can be identified prior to the transition: in the first step crystallization 
involves an increase of orientational order even in the liquid state, only later followed by a change of density (positional order).
This decoupling can be intuitively understood on the basis that liquids are difficult to compress, and some order
has to appear for realizing efficient packing before density can increase. 

In this Article, we will consider several examples of 
two-step crystallization.  
The term ``two-step'' was originally applied to crystallization processes which involved an intermediate step corresponding to metastable phases
where nucleation occurs at a higher rate. It has then been extended to describe crystallization processes that are continuous,
but where nucleation occurs in preordered regions of the melt.
We will start with considering how a metastable
gas-liquid phase separation alters the nucleation pathway by adding an intermediate step, i.e. the formation of a dense liquid
droplet, whose high density considerably increases the crystallization rate. We will then consider examples where
the nucleation pathway occurs through
crystallization precursors spontaneously formed in melts. In this case we show that the liquid-to-solid transition is effectively driven
by structural fluctuations, and the nuclei reach their bulk density only when their size considerably exceeds the critical
nucleus size. In the case of spherically symmetric interaction potentials, we show that crystallization occurs in regions
of high crystal-like bond-orientational order, and that these regions effectively select the polymorph that will be nucleated from them. 
Connection are made also with vitrification phenomena for hard spheres, in which we associate crystalline orientational order with slow regions in the system, and 
identify five-fold symmetric regions as inhibitors of crystallization. It is worth stressing that such bond orientational ordering does not accompany the density change.
We will then focus on the special case of water, where the presence of directional hydrogen bonds considerably alters the pathway
to nucleation.

All these indicate that a liquid is not a homogeneous isotropic state at the microscopic level, but generally tends to have more local structural order to lower its free energy. 
This tendency, which increases for a lower temperature, has an impact on the structural and dynamical properties of a supercooled liquid state, which further influences crystal nucleation and vitrification \cite{tanaka2012bond,tanaka2013importance}. 
In this Article, we consider only a (nearly) single-component system, where density and bond orientational order are two important order parameters 
to describe the liquid state. However, for a multi-component system, we need an additional order parameter, local composition, to describe it. 
Here we do not discuss this important but difficult case. 
We are going to show that the multiple-order-parameter description is crucial to understand the low-temperature phenomena seen in a supercooled liquid state.

\section{Classical Nucleation Theory}

The simplest and most general understanding of nucleation is embodied in Classical Nucleation Theory (CNT)~\cite{becker1935kinetische,zeldovich1943theory,kelton2010nucleation}.
Focusing here on the liquid-to-solid transition, the process starts from small crystalline nuclei that spontaneously form
in the supercooled liquid from thermal fluctuations. A critical nucleus size, at which these nuclei are equally likely to dissolve or to grow, can be easily obtained from the Gibbs-Thomson equation.
The chemical potential of a solid particle inside a nucleus of size $R$ is given by
$$
\mu_s(r)=\mu_s^0+\frac{2\gamma}{\rho_s R},
$$
where $\mu_s^0$ is the bulk chemical potential of the solid phase, $\gamma$ is the interfacial tension,
and $\rho_s$ is the bulk density of the solid. At the critical nucleus size $R_c$, the crystalline nucleus is in 
equilibrium with the liquid phase, $\mu_s(r_c)\equiv\mu_l^0$, from which we obtain
$$
R_c=-\frac{2\gamma}{\rho(\mu_s^0-\mu_l^0)}.
$$

Below the critical nucleus size, the clusters made of $n$ particles are thermally equilibrated at all times, meaning that they are evolving
in a free energy landscape where the nucleation rate density can be written as 
\begin{equation}
\kappa=K\exp(-\beta\Delta F(n_c)),
\end{equation}
where $K$ is a kinetic factor, and $\Delta F(n_c)$ is the height of the free energy barrier separating the
liquid from the solid phase. Here $n_c$ is the number of particles in the critical nucleus.
According to CNT, the kinetic term takes the form
\begin{equation}\label{eqn:k}
K=(\rho/m) f_n^*\sqrt{\beta\Delta F^{''}(n_c)/2\pi}, 
\end{equation}
where $\rho$ is the density, $m$ is the molecular mass, $f_n^*$ is the attachment rate of particles
to the critical nucleus, and $F^{''}(n_c)$ is the curvature of the barrier at
the critical nucleus size $n_c$. The term $Z=\sqrt{\beta\Delta F^{''}(n_c)/2\pi}$ is
also known as the Zeldovich factor.
The free energy barrier $F(n)$ is instead derived from a competition between a bulk term
which accounts for the free energy gain of transforming to the stable phase, and a surface
term which expresses the free energy penalty in creating a liquid/solid interface
\begin{equation}
\Delta F(n)=n\Delta\mu+c \gamma n^{2/3},
\end{equation}
where $c$ is a numerical factor that depends on the shape of the nucleus ($c=36\pi/\rho^2$ for a spherical nucleus), 
$\Delta \mu$ is the liquid-solid chemical potential difference, and $\gamma$ is the liquid-crystal surface tension. 

Here it is worth mentioning that the transport crucial for crystal nucleation and growth 
is not controlled by viscosity, but by translational diffusion (see, e.g., Ref. \cite{tanaka2003possible}). This is important below the melting point, where 
the Stokes-Einstein relation rather significantly breaks down.

In the case of heterogeneous nucleation, the free energy barrier is decreased by a geometrical factor
$$
\Delta F_\text{het}=\Delta F_\text{hom}\frac{(1-\cos\theta)^2(2+\cos\theta)}{4}, 
$$
where $\theta$ is the contact angle of the crystal nucleus with the external surface.

\section{Extending Classical Nucleation Theory}\label{sec:extending}
Classical Nucleation Theory provides a successful framework to understand and analyze data from nucleation
for a large variety of processes. But, on quantitative level, its predictions are hard to test, as small
changes in the experimental conditions can significantly impact the yield rates of crystallization.
In the introduction section, we highlighted the large discrepancies in the nucleation rates obtained experimentally
and theoretically for arguably the simplest system that exhibits a liquid-to-solid transition, i.e. colloidal hard spheres.
Several attempts have been made at solving this inconsistency, but no consensus has been found yet.
Approximations in the calculations of nucleation rates can be found both for experiments and simulations,
and explanations are being proposed which take into account the limitations of both methods. 
In one recent proposal~\cite{russo2013interplay}, the effects of the gravitational fields on the experimental
nucleation rates are considered. The difficulty of exactly density
matching colloidal samples implies that gravitational lengths are not negligibly short compared to the size of
colloidal systems. The coupling between the gravitational force with long range hydrodynamic effects can then
cause an increase of density fluctuations in the sample, followed by an increase of nucleation rates compared
to the gravity-free case.
A different proposal~\cite{radu2014solvent} instead focuses directly on hydrodynamic interactions, which are often
neglected in simulations. It is argued that hydrodynamic interactions can effectively increase nucleation rates
to account for the discrepancy between experiments and simulations.
To this day, consensus on the resolution of the discrepancy in nucleation rates has not been found.

Classical Nucleation Theory rests on several approximations.
The most important one is the so-called
\emph{capillarity approximation}, which is the assumption that small crystalline nuclei
(or the order of 100 of particles) are still amenable to a thermodynamic description, as they retain the same
properties of the bulk solid.
This assumption has been challenged in many studies~\cite{yau2001direct,yau2000quasi,strey1993measurement,oxtoby1998nucleation,granasy1997comparison}.

\begin{figure}
 \centering
 \includegraphics[width=8cm]{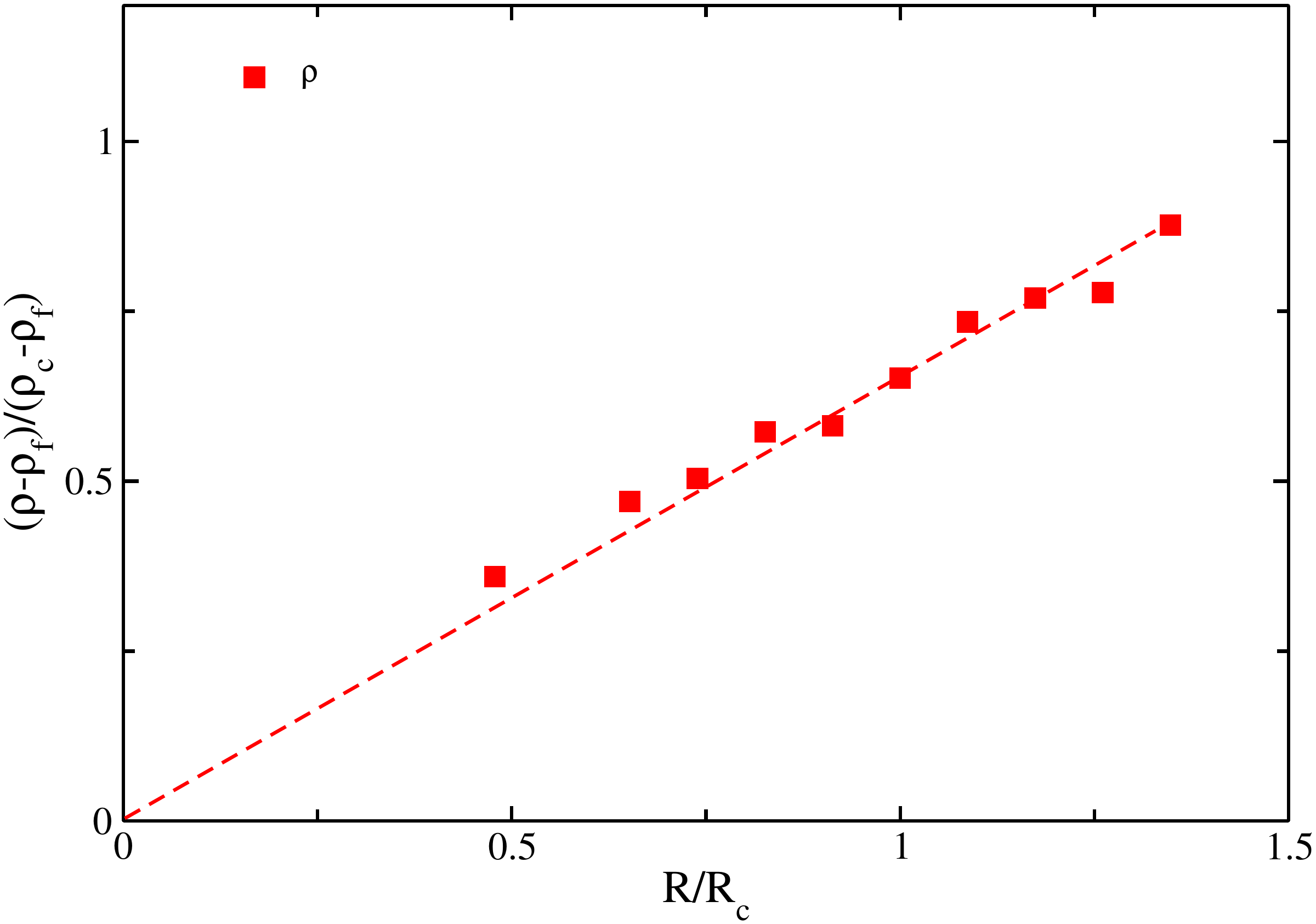}
 \caption{Density of crystalline nuclei as a function of their size. Crystalline nuclei for the hard-sphere system
 are spontaneously formed at reduced pressure $\beta P\sigma^3=17$, and the average density at the center of mass
 is plotted against the radius $R$ of the nucleus. $\rho_f$ and $\rho_s$ are the equilibrium densities of the fluid
 and of the crystal respectively, while $R_c$ is the critical nucleus radius.}
 \label{fig:density_nucleus}
\end{figure}

As an example we consider here a hard sphere system at reduced pressure $\beta P\sigma^3=17$, where $\beta=1/k_BT$, and $\sigma$ is the diameter of the spheres 
(see Ref. \cite{russo_hs}). At these conditions the critical nucleus size consists approximately of $n_c=85$ spheres, and direct crystallization
events can be observed in simulations. By gathering statistics on many independent crystallization events, in Fig.~\ref{fig:density_nucleus} we plot the
average density at the center of mass of the nucleus as a function if its radius $R$. Density is normalized with the density of the fluid $\rho_l$ and of
the solid $\rho_s$ at their equilibrium values, so that it is $1$
in the bulk $fcc$ solid, and $0$ in the metastable liquid phase.
The radius is instead scaled with the critical radius $R_c$. From the figure we can see
that at equilibrium, $R/R_c=1$, the density at the center of the nucleus is far below the equilibrium density of the crystal $\rho_c$.

Another parameter that is not well described by the capillarity approximation is the surface tension $\gamma$. Several studies have found that
the surface free energy cost at equilibrium is not the relevant free energy penalty for nucleation, both in metals and charged colloids~\cite{jiang2008size,bokeloh2011nucleation}.

Among the many theoretical treatments that go beyond the capillary approximation see for example Ref.~\cite{oxtoby1998nucleation,granasy1997comparison,prestipino2012systematic}:
these include modifications of CNT (as for example introducing a lenghtscale-dependent surface tension, i.e. Tolman length)~\cite{prestipino2012systematic},
and new applications like density functional theory~\cite{harrowell1984molecular,bagdassarian1994crystal,shen1996nucleation,oxtoby1998nucleation,lutsko2011communication,turci2014solid} and kinetic nucleation theory~\cite{girshick1990kinetic}.

It is worth noting that in most descriptions of nucleation, the melt is described as structure-less and homogeneous.
Several studies have shown instead that the supercooled state contains a lot of structural order aside from crystalline order~\cite{tanaka2012bond,tanaka2013importance}.
For hard spheres, for example, the competition between crystalline order and icosahedral (or five-fold symmetric) order plays
a big role in the nucleation process, and on its avoidance~\cite{leocmach2012roles,kawasaki,kawasaki2010correction,russo_hs,mathieu_russo_tanaka,karayiannis2011fivefold,laso_soft,royall2015role}.
For systems with a metastable critical point, e.g. with short-range attractions, spatial heterogeneity due to critical fluctuations also plays a big role~\cite{ten1997enhancement}.

CNT rests on the assumption that the nucleation process can be described by just one reaction coordinate,
and that all order parameters involved
in the transition proceed simultaneously. Several works have addressed this assumption in systems where different
reaction coordinates can be followed simultaneously during nucleation, and have demonstrated that
a single-order parameter description fails at capturing important aspects of the transition~\cite{kawasaki,kawasaki2010correction,pan2004dynamics,moroni2005interplay,trudu2006freezing,peters2006obtaining,zykova2008irreducible,lechner2011role,russo_hs,
prestipino2014shape,lutsko2015two}. 
In the following we will focus on this particular aspect of the nucleation process, and show how new insights can
be obtained with a multi-dimensional (or, multi-order-parameter) representation of the transition.

In colloidal systems, unlike metals and noble gases, polydispersity plays a big role in the crystallization
process. Formally speaking, a continuous polydispersity introduces a new order parameter for each specie in the mixture~\cite{fasolo2003equilibrium,sollich2011polydispersity}.
At high polydispersities, crystallization requires fractionation, and generally destabilizes the crystal. At low polydispersities, it is
possible to treat the system like an effective one-component system, but care has to be taken in describing the nucleation process, as several thermodynamic
quantities are very sensitive to polydispersity, such as surface tension~\cite{palberg2016equilibrium}.

It is important to note that kinetic effects can have a profound impact on the crystallization process. Kinetic effects become
especially important at high supercooling, where diffusion becomes the rate limiting process for both nucleation and crystal growth.
According to Classical Nucleation Theory, the kinetic term in Eq.~(\ref{eqn:k}) is controlled by a thermally activated process~\cite{kelton2013crystal},
usually thermal diffusion (at low temperatures diffusivity and viscosity can have different activation energies, i.e. Stokes-Einstein violation) 
\begin{equation}\label{eqn:k2}
K=Z\rho\frac{24Dn_c^{2/3}}{\lambda^2} 
\end{equation}
where $\lambda$ is the atomic jump distance in the liquid.
In the crystallization of a Lennard-Jones system~\cite{ten1996numerical}, it was shown that
the crystallization rate predicted by Eq.~(\ref{eqn:k2}) is two orders of magnitude smaller than
the measured one.
This suggests that, for simple monoatomic systems, the kinetic factor is not controlled by thermally activated processes.
This behaviour is consistent with the growth of the Lennard-Jones crystal~\cite{broughton1982crystallization} where the crystallization rates
were found to be independent of diffusion, but determined by the ideal gas thermal velocity $(3k_BT/m)^{1/2}$.
This means that, at high supercooling, crystal growth can proceed despite diffusive processes becoming negligible, even below the glass transition temperature.
In a series of experiments~\cite{hatase,konishi,roy2} on molecular liquids (both organic and inorganic
glass formers), it was discovered that the crystal growth rate can display
an unusual enhancement below the glass transition temperature. The growth rate
becomes orders of magnitude faster than expected for diffusion-controlled growth,
in conflict with the established view that the kinetic barrier for crystallization
is similar to that for diffusion. Simulations of hard spheres have also shown
crystallization without diffusion~\cite{zaccarelli},
that the static properties of the crystals are not much affected by dynamics~\cite{valeriani2012compact},
and that the dynamics in a crystallizing glass becomes intermittent (avalanche like)~\cite{sanz2014avalanches}.
There have been several attempts at describing the process by which particles move from the liquid to the crystal
without assuming an activated process analogous to that which governs self-diffusion.
These include the substitution of diffusion with
secondary relaxations~\cite{hatase} (the so called $\beta$ processes), or by taking into
account the extensional stress around crystals growing inside a glass of
lower density~\cite{tanaka2003possible,konishi,caroli2012ultrafast}. 
Despite these attempts, a clear explanation is still lacking. For example, simulations of
binary metallic glasses have instead confirmed diffusional growth of the interface~\cite{tang2013anomalously}.
These simulations have noticed a connection between the attachment rate of liquid particles
to the crystal surface with the width of the surface, which suggests that the structural properties of the
melt, and its preordering in advance of the front propagation, might play a bigger role than dynamics.
We will investigate the role of structural preordering of the fluid in detail in Section~\ref{sec:precursors}.

Despite its limitations, CNT is still our best effective theory of the nucleation process.
Rather than focusing on the underlying symmetries (translational and orientational order)
CNT uses the size of the nucleus as its reaction coordinate, and gives a unified description
of nucleation in a big variety of systems.
But the simplifications that are adopted (capillarity approximation, and reduction to a one-dimensional description)
hide a lot of the phenomenology of the nucleation process, and in the next sections we unveil some of the steps
that have been made in understanding nucleation from a microscopic perspective.

\section{Two-step nucleation: Coupling between density fluctuations and crystal ordering}\label{sec:two-step}
One of the first demonstrations of a non-classical pathway to nucleation was given by ten Wolde and Frenkel~\cite{ten1997enhancement}.
They considered how the crystallization pathway is influenced by the presence of a metastable gas-liquid transition, whose order parameter 
is a scalar density field.
The model considered was an adapted Lennard-Jones potential, with a hard-repulsion at contact, and short-range attraction.
In Soft Matter systems, reducing the range of the attractions pushes the gas-liquid critical point to lower temperatures,
where it eventually becomes metastable with respect to crystallization~\cite{anderson2002insights}. The parameters were chosen in order to have a critical point
lying approximately 20\% below the equilibrium crystallization curve, a condition that is found to be common in globular protein solutions.
According to the study, a novel crystallization pathway was found in proximity of the gas-liquid critical point. Instead of direct nucleation
from the gas phase (one-step crystallization), the system was found to crystallize inside disordered liquid regions of higher density that form due to
critical density fluctuations. This process is referred as two-step crystallization: the first step involves
the formation of dense liquid regions, and the second step is the nucleation of the crystal phase inside these dense regions. 
In the original work~\cite{ten1997enhancement} the formation of dense precursors was attributed to
the long-range density fluctuations that originate from the presence of the metastable critical point. It is thus expected
that the nucleation rate would have maximum around the critical point. A recent study~\cite{wedekind2015optimization} has instead
found that that the nucleation rate monotonically increases going from the supercritical region towards
the two-phase coexistence region. The two-step mechanism is thus not strictly due to critical fluctuations, but rather to the formation of a dense liquid phase that is
thermodynamically stabilized below the critical point.
Away from the metastable critical point, the system was found to crystallize classically in one step, where densification and structural
ordering happen simultaneously. Thus it was demonstrated how a very simple model (spheres with short range attractions) can display very complex crystallization behavior depending on the thermodynamic parameters~\cite{haxton2015crystallization}.

In experiments, two-step pathways have been first considered in protein crystallization~\cite{thomson1987binary,georgalis1997lysozyme,muschol1997liquid,igarashi1999initial,sauter2015question}, most notably lysozyme, but were quickly observed also in colloidal systems~\cite{zhang2007does,savage2009experimental}.
Numerous examples of two-step crystallization from solutions has been reported in the last years, and several reviews have been
written on the subject~\cite{erdemir2009nucleation,vekilov2010two,gebauer2014pre}.

Probably the best evidence for two-step nucleation pathways came from theoretical investigations, 
where microscopic information is readily available. The main theoretical tool for the study
of crystallization in simple model systems are density functional theory (DFT)~\cite{talanquer1998crystal,lutsko} and molecular simulations~\textcolor{blue}{\cite{asherie1996phase,soga1999metastable,sosso2016crystal}}.

Two-step crystallization pathways have been suggested to occur also outside the region of stability of the dense fluid phase.
In this case, crystallization starts inside metastable liquid drops (also called pre-nucleation clusters~\cite{gebauer2014pre})
whose lifetime has to be long enough to promote nucleation.
The first example came from experimental studies of lysozyme crystals, which were found to crystallize more easily just outside the
metastable gas-liquid phase separation transition~\cite{galkin2000control,vekilov2004dense}. Similar processes have been found in 
both small molecules, like glycine~\cite{chattopadhyay2005saxs} and calcium carbonate~\cite{gebauer2014pre}, and colloidal systems~\cite{savage2009experimental}.
Two-step crystallization in absence of a metastable liquid phase was also confirmed theoretically in spin models~\cite{sear2009nucleation} and
density functional theory~\cite{lutsko}.

Even the presence of metastable liquid phases might not be a necessary condition, as several studies have reported two-step nucleation
pathways in systems where a dense fluid phase does not exist. Hard spheres, due to the lack of attractive interactions, are an example of such systems: 
the stable phases are a fluid phase and close-packed crystalline phases, such as hcp, fcc and the phase obtained
by random stacking of hcp and fcc layers. The large size of colloidal particles means that the lengthscales and timescales can
be (relatively) easily observed experimentally, and many classical experiments have examined the process of crystal nucleation
in microscopic detail~\cite{pusey,zhu1997crystallization,gasser,martelozzo2002structural}. Also simulations have studied in detail
the composition, shape and free energy barrier to nucleation~\cite{auer2001prediction,bolhuis,filion,filion2}. 
Leveraging the sensibility of light scattering experiments to density fluctuations, dense 
amorphous precursors were reported in colloidal hard spheres~\cite{schope2006two}, PMMA particles~\cite{schope2007preparation,schope2006small,schope2007effect}, and microgel colloids~\cite{iacopini,franke2014solidification}. These experiments were also corroborated by molecular dynamics simulations~\cite{schilling,schilling_jpcm}.

Whether dense precursors can be formed in systems without a metastable fluid-fluid demixing transition (like hard spheres) is not so clear physically and has been debated in the literature.
In Ref.~\cite{russo_hs} we argued that orientational order foreshadows translational order in hard-spheres, i.e. that the nucleation process is accompanied
by a faster increase of bond-orientational order than translational order. This also implies that precursor regions, defined as the regions of the melt
from which the critical nucleus will emerge from, are better characterized by bond orientational order than density.
In the next Section we will go through the different arguments in favor of this scenario, and in Section~\ref{sec:whosfirst} we will comment briefly over the comparison with other scenarios.

\section{Nucleation precursors}\label{sec:precursors}

\subsection{Crystallization of liquids interacting with isotropic potentials}
\subsubsection{Roles of precursors in the birth of crystals}
\begin{figure}[!t]
 \centering
 \includegraphics[width=8.5cm]{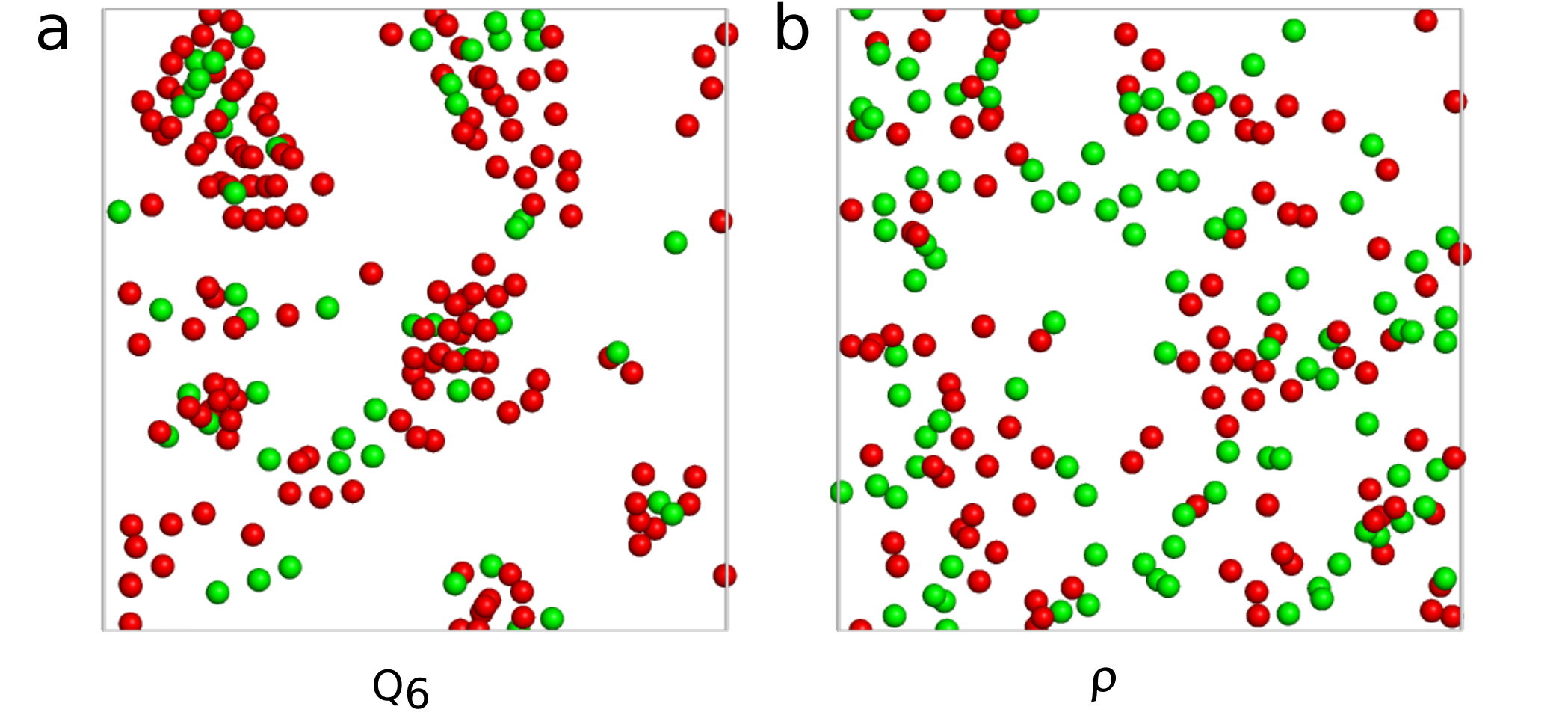}
 \caption{$Q_6$ and $\rho$ fluctuations in hard spheres at reduced pressure $\beta P\sigma^3=17$.
(a) Snapshot of particles with the $10\%$ highest value of $Q_6$. Red particles have higher density than the median
 density; green particles have lower density than the median density.
(b) Snapshot of particles with the $10\%$ highest value of $\rho$. Red particles have higher $Q_6$ than the median
 $Q_6$; green particles have lower $Q_6$ than the median $Q_6$. Adapted from Ref. \cite{russo_proceeding}.
 }
 \label{fig:fluctuations}
\end{figure}

In the previous section we focused on pathways where the metastability of a liquid phase induces  
the formation of dense drops that promotes crystallization. We will show here that, in
absence of a metastable gas-liquid transition, the opposite scenario is instead favored,
with structural order fluctuations preceding the increase of density. 
This can be intuitively understood on the basis that the melt is difficult to compress, and some structural order has to appear 
due to packing effects before density can increase.

In order to study this process from a microscopic perspective, one should follow the nucleation process from the nucleation event
to the growth of the crystalline nucleus over its critical nucleus size. By using local order parameters, describing the degree of
translational and orientational order around individual particles, it is then possible to construct a statistical map of the microscopic
pathway to crystallization in a multiple-order-parameter space. While most of our knowledge comes from molecular simulations on model systems~\cite{kawasaki,kawasaki2010correction,kawasaki2010structural,schilling,russo_hs,russo_gcm,russo2013interplay}, recent
confocal microscope experiments have also started to shed light on this process in colloidal systems~\cite{tan2014visualizing,lu2015experimental}.

Translational order can be computed from two-body correlation functions, such as the pair distribution function or the
two-body excess entropy~\cite{russo_hs,mathieu_russo_tanaka}, but it is found that all these measures correlate well
with the local density (as obtained from the Voronoi diagram of particle configurations). It is thus common to use the local
density as a measure of translational order. 
Unlike translational order, which is obtained
from two-body correlation functions, orientational order is obtained by considering many-body correlations.
For spherical particles, an adequate measure of orientational order was introduced by
Steinhardt et al.~\cite{steinhardt}, and later popularized by Frenkel and co-workers in the study of crystallization~\cite{auer2001prediction,auer2004quantitative}.
The order parameter most often used for hard spheres is $q_6$, where the subscript $6$ denotes the use of spherical harmonics of degree $6$,
and it is computed for each particle in the system by considering only the position of nearest neighbors ($\sim 12$ for hard spheres).
Lechner and Dellago have also proposed to coarse-grain orientational order parameters, significantly reducing fluctuations
by including the orientational order up to the second shell of nearest neighbors~\cite{lechner}. We use here capital letter to denote coarse
grained versions of orientational order parameters: for example, $Q_6$ is the coarse-grained version of $q_6$.
For the exact definitions of these order parameters we refer to Ref.~\cite{lechner,russo_hs}. 
Here we note that a combination of more than two bond orientational orders is usually necessarily to specify a particular rotational symmetry. 
For details, please refer to Ref. \cite{leocmach2012roles,russo_hs}.

In the following we focus on the hard-sphere system, as a model of fluids that crystallize in close-packed structures.
We compare the bond orientational field and the density field according to five different criteria.

 \begin{figure}[!t]
 \centering
 \includegraphics[width=8cm]{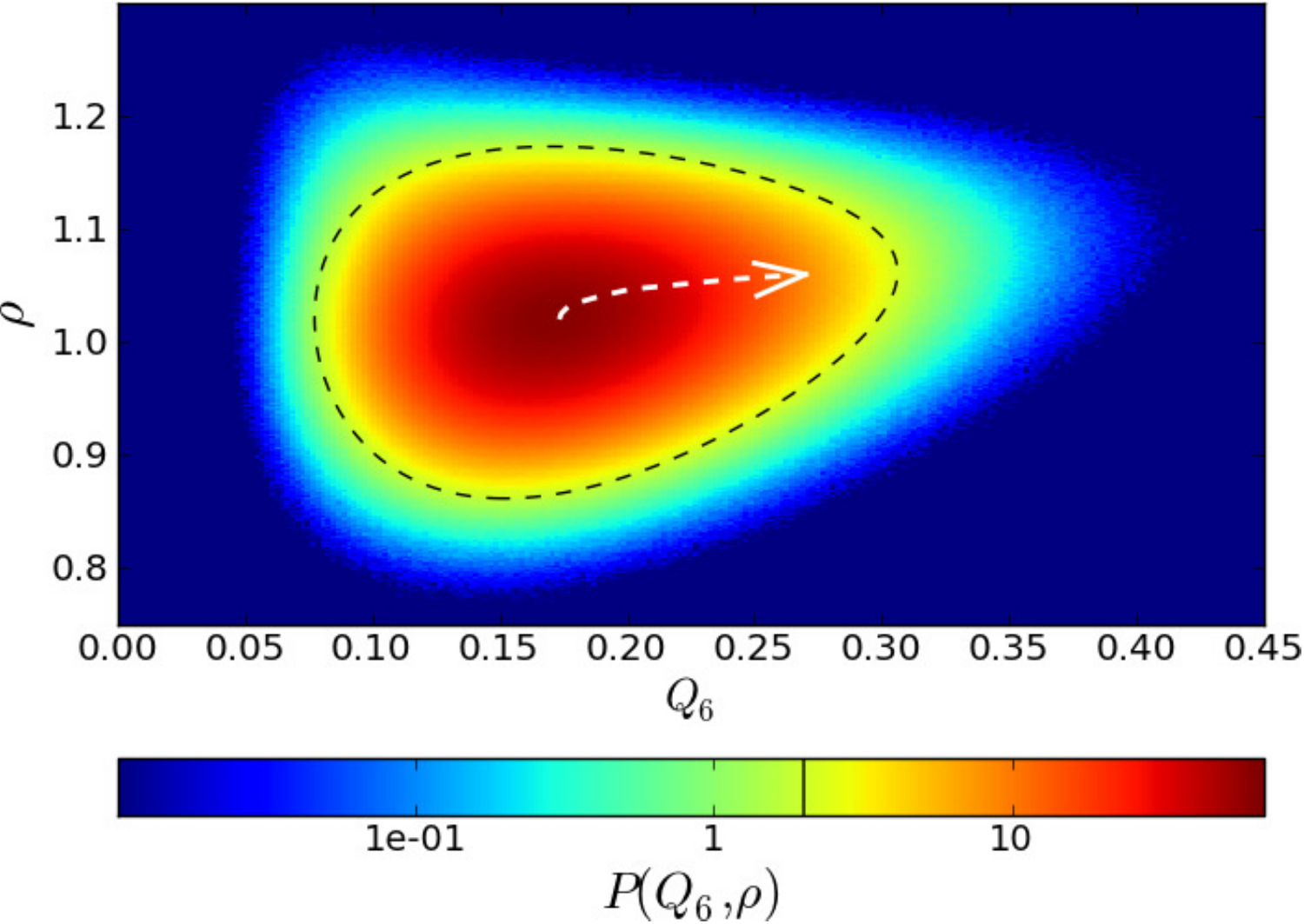}
 \caption{Landau free energy. Joint probability distribution function $P(Q_6,\rho)$ for the supercooled melt at $\beta P\sigma^3=17$. The dashed white arrow is the steepest descent path
from the maximum of the distribution. Adapted from Ref. \cite{russo_proceeding}.}
 \label{fig:map}
\end{figure}

\begin{description}
 \item[Correlation length of the fluctuations] In a two-step scenario, one expects
 the crystalline nucleus to appear inside a region where the relevant order parameter is higher than the rest of the melt.
 The size of this region should also increase with supercooling. 
 It is well known that the length-scale of density fluctuations barely changes with supercooling, while
 Refs.~\cite{ShintaniNP,kawasaki2007correlation,tanaka,kawasaki2010structural,mathieu_russo_tanaka} showed indeed that the correlation length of bond orientational order increases with supercooling.
 For example, Figs.~\ref{fig:fluctuations}(a) and (b) show that while density fluctuations are almost uncorrelated, bond orientational order is much more correlated. Direct
 observation of nucleation events have shown that the nuclei do appear to form inside regions of high bond orientational order~\cite{kawasaki,kawasaki2010correction,kawasaki2010structural,russo_hs}.
 \item[Lifetime of fluctuations] In order to promote crystallization, the fluctuations need to be sustained enough for a nucleus
 to appear inside them and eventually reach the critical size. The lifetime of density fluctuations is given by the decay of the intermediate
 scattering function, which is used as a definition of the structural relaxation time, $\tau_\alpha$. Several studies have shown that structural
 fluctuations can last much longer than $\tau_\alpha$~\cite{kawasaki,kawasaki2010correction,ShintaniNP,kawasaki2007correlation,tanaka,kawasaki2010structural,malins2013identification,malins2013lifetimes,royall2015role}. 
 This is particularly the case for two-dimensional liquids \cite{tanaka}.
 \item[Landau free energy] The Landau free energy in the two-dimensional space of bond orientational order and density \cite{tanaka2012bond,tanaka2013importance} is
 obtained by computing the joint probability of fluctuations in $Q_6$ and $\rho$, $F(Q_6,\rho)=-k_BT\log P(Q_6,\rho)$. 
 Figure~\ref{fig:map}
 shows the result for hard spheres at reduced pressure $\beta P\sigma^3=17$, taken from Ref.~\cite{russo_proceeding},
 and displays a good decoupling
 between $Q_6$ and $\rho$, expressing the fact that $Q_6$ and $\rho$ capture independently the fluctuations
 in bond orientational order and density~\cite{russo_proceeding}. A cubic fit to the free energy shows that the dominant cubic term is of the form $Q_6\rho^2$ 
 (see Ref. \cite{tanaka2012bond,tanaka2013importance} for a possible form of the free energy functional). Because
 the interaction is quadratic in $\rho$ and linear in $Q_6$, the system can increase its orientational order without increasing the translational order, 
 but the contrary is not true. This constrains the fluctuations towards a stronger increase in its orientational order. The analysis also shows a
 weak linear coupling between $\rho$ and $Q_6$, which is visible in the small positive tilt of the free energy contour in Fig.~\ref{fig:map}.
 This linear term also indicates that regions of high orientational order will, on average, have also higher density than the melt (see Fig. \ref{fig:fluctuations}(a)). But the opposite
 is not true: due to the stronger $Q_6\rho^2$ term in the free energy expansion, regions of high density are not, on average, characterized by higher $Q_6$ (see Fig. \ref{fig:fluctuations}(b)).

 \begin{figure}[!t]
 \centering
 \includegraphics[width=8.5cm]{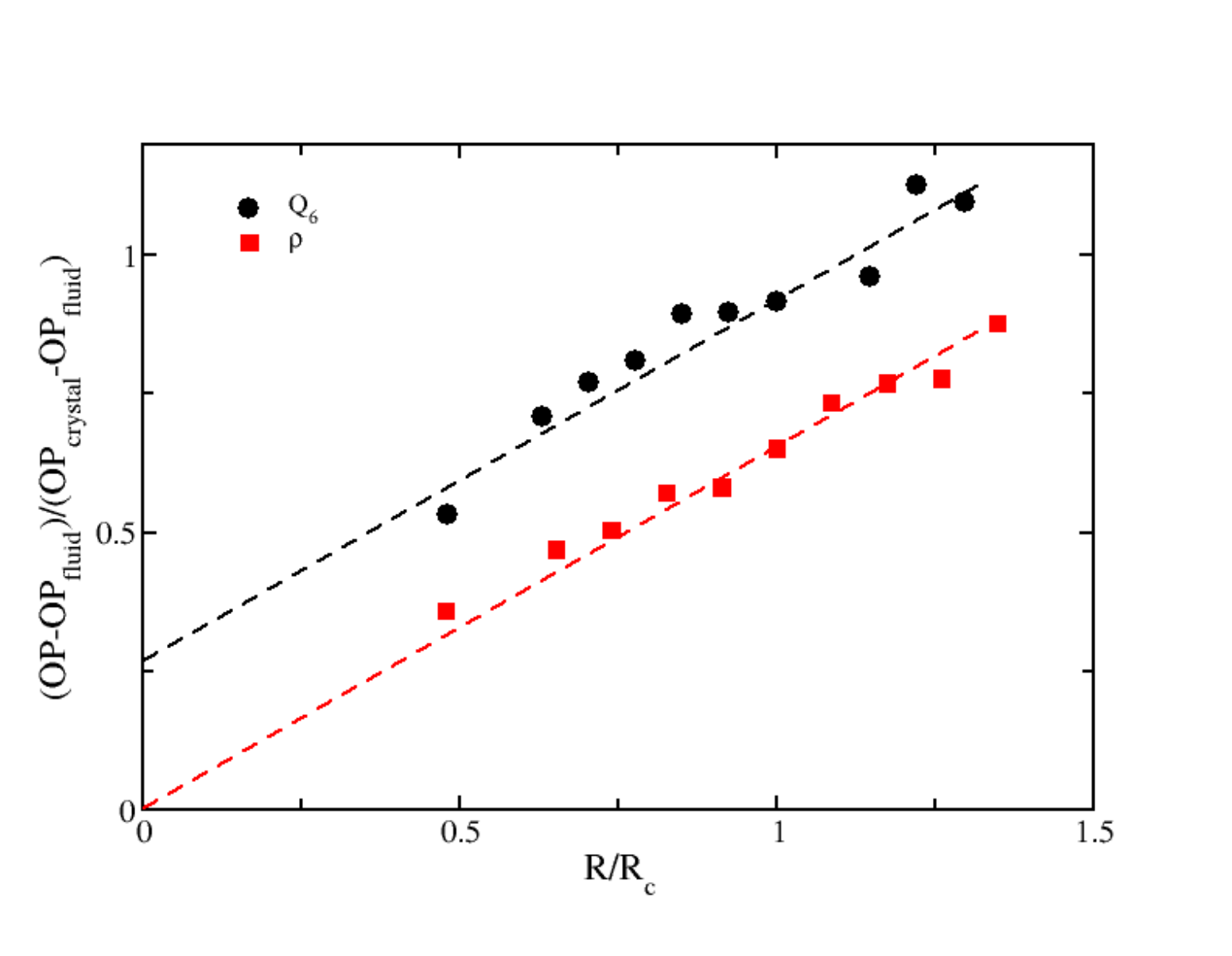}
 \caption{Density and bond-orientational order of crystalline nuclei as a function of their size. Average values of the order parameters (density in red squares, $Q_6$ in black circles) at the center of a crystalline nucleus (its center of mass) as a function of  the scaled nucleus radius, $R/R_c$, where $R_c=85$ is the critical nucleus size.}
 \label{fig:radial}
\end{figure}
 \item[Radial profile of crystalline nuclei] It is possible to compute the average radial profile for both $\rho$ and $Q_6$ in crystal nuclei of a certain size. The results show that,
 going from the fluid phase towards the center of the nucleus, the system first develops orientational order, only later followed by an increase of density~\cite{russo_proceeding}. 
 Similar results have been obtained with density functional theory~\cite{oettel2012mode,turci2014solid}.
 At the critical nucleus size, the center of the nucleus is closer to the $Q_6$ of the bulk crystal, yet the density still being just $60\%$ of the bulk value
 for a critical nucleus of size between 80-90 particles. Extracting the values of the order parameters at the center of the nucleus as a function of
 nucleus size, Fig.~\ref{fig:radial} shows that the density extrapolates to the fluid density at zero size, while $Q_6$ seems to form at values higher than 
 the bulk fluid. This clearly suggests that indeed the nuclei appear from regions of high bond orientational order, but whose density is still comparable
 to the fluid's density. This means that crystal nucleation takes place heterogeneously in space. 
 \begin{figure}[!t]
 \centering
 \includegraphics[width=7cm]{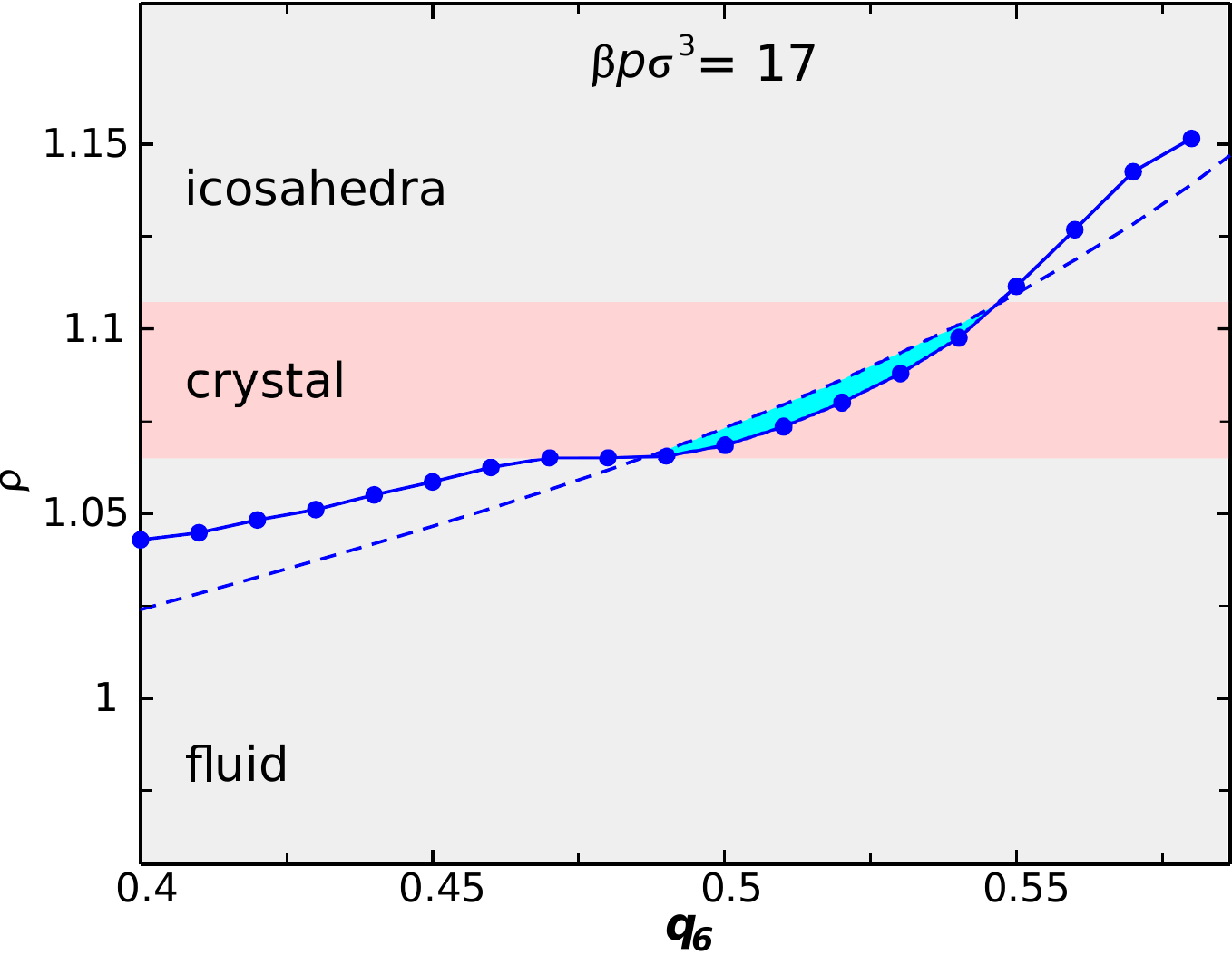}
 \caption{Translational vs Orientational order curves. Average density as a function of the local $q_6$ for crystalline particles (symbols and solid line) and non-crystalline particles (dashed line) 
in a metastable liquid state at $\beta P \sigma^3=17$.}
 \label{fig:curves}
\end{figure}
 \item[Translational vs Orientational order curves] Ref.~\cite{russo_hs} introduced the idea of plotting the average density for a metastable liquid state as a function of orientational
 order for different subsets of particles, as for example liquid and solid particles. The figure is reproduced in Fig.~\ref{fig:curves}, showing that: i) the solid branch shows a characteristic plateau in $\rho$ after which it becomes the stable branch at that state point;
 ii) the solid branch looses stability at volume fractions of approximately $58\%$, thus close to the glass transition. The presence of the distinct plateau in $\rho$ clearly shows that
 the transition occurs at constant density and involves instead an increase of bond orientational order.
 Ref.~\cite{russo_hs} showed that the liquid and solid branch 
 start crossing approximately at the melting pressure, below which the two curves do not cross.
 Similar calculations have shown how polydispersity
 destabilizes the solid branch~\cite{mathieu_russo_tanaka}, and have been found to apply in colloidal experiments~\cite{tan2014visualizing,lu2015experimental}.
\end{description}

Evidence of bond orientational foreshadowing of crystallization has been suggested for colloidal systems both in simulations~\cite{kawasaki,kawasaki2010correction,kawasaki2010structural,russo_hs,russo_gcm,mathieu_russo_tanaka,kratzer2015two} and  experiments~\cite{tan2014visualizing,lu2015experimental,mohanty2015multiple}.
Interestingly, the role of structural fluctuations has been investigated also in a variety of different systems, like metallic melts~\cite{kelton2013crystal,li2014nucleation,debela2014nucleation}, anisotropic particles~\cite{han2013shape,mahynski2014stabilizing,mondal2015glass},
and polymers~\cite{karayiannis2012spontaneous,hoy2013simple}. Charged colloids are good model systems for metals and alloys, and a parallel between the two systems,
in which the development of short-range order is highlighted, can be found in Ref.~\cite{herlach2010colloids,herlach2014colloids,herlach2016experimental}.
Recent advances in X-ray spectroscopy~\cite{lehmkuhler2014detecting,schroer2015nano,lehmkuhler2016intramolecular,liu2016calculation} hold the potential of computing bond orientational order from scattering experiments,
and could open a new era in the investigation of local structural information in disordered systems.
The idea of studying crystallization under external fields by studying the
effects of the field on the bond orientational order is also a new direction for future research, which has been considered for example
in Ref.~\cite{lander2013crystallization} for the case of crystallization in sheared suspensions.

\subsubsection{``bond-order-first'' vs ``density-first''}\label{sec:whosfirst}

There has been some debate in the literature as to whether the nuclei appear in dense-precursors~\cite{schilling}
or in bond-orientational-ordered precursors~\cite{kawasaki,kawasaki2010correction,russo_hs}. A very recent~\cite{berryman2016early} re-analysis of the simulation trajectories
of Ref.~\cite{schilling} (``density-first'' case) showed a simultaneous increase of density
and bond-orientational order leading up to nucleation. 
As we observed in Fig.~\ref{fig:map} there is a (weak) linear coupling between bond-orientational order and density which ensures that fluctuations toward high bond-orientational order will on average lead to a higher density than the melt. This also means, as found in Ref.~\cite{berryman2016early}, that following crystal nucleation in time will lead to a simultaneous increase of density and bond-orientational order,
as was observed in Ref.~\cite{russo_hs}.

This consideration may also have an implication on light scattering results of crystal nucleation 
\cite{franke2014solidification}. 
Light scattering measurements are basically sensitive to the development of translational order, but not to that 
of bond orientational order. Thus, the nucleation process
can only be followed as soon as the scattering from the [111] plane is observed, but still before the formation of any higher order
peaks. At these resolutions, the distance between [111] planes is observed to be considerably shorter than the nearest neighbour spacing in the remaining melt, which signals that densification is also occurring during the formation of 
2D sheets of hexagonal order.

The fact that both order parameters are changing during the transition, does not mean that their role is the same.
Positional order and orientational order are characterized by their spatial correlation functions. It has been shown~\cite{tanaka,kawasaki,kawasaki2010structural,leocmach2012roles,mathieu_russo_tanaka}, in both mono- and poly-disperse systems, that with supercooling
(compression over the freezing pressure) hard-spheres rapidly develop orientational order with negligible signs of increase in translational order:
while the correlation length of positional correlation stays constant with supercooling, the bond-orientational correlation length steadily increases, almost
doubling its size with respect to fluid state below the freezing pressure.
So there is a clear decoupling between orientational and positional order, even if it is not strong enough to sustain an intermediate phase between the fluid and crystal
(analogous to the hexatic phase in two-dimensional hard-disks~\cite{bernard2011two}). Instead, the simple first-order nature of the fluid-to-solid
transition, dictates a discontinuous change of positional and orientational order, and instead of two-step crystallization (like what happens near a metastable
critical point) it is more appropriate to speak of precursors mediated crystallization in hard sphere like systems.

As we have observed in Fig.~\ref{fig:map}, the two order parameters do not have symmetric roles in the free energy, and while higher bond-orientational order
corresponds to higher density, the opposite is not true. As we discussed in Section~\ref{sec:two-step} (Fig.~\ref{fig:map}), the linear coupling between the two
order parameter is weaker than the term $Q_6\rho^2$, and higher density does not imply higher bond-orientational order.
In fact, there are many local configurations which increase the density locally, but do not increase crystalline bond-orientational order. The most notable
of these structures are icosahedral clusters, which correspond to the highest local packing of hard-spheres, but that are incompatible with
crystalline bond-orientational order. While at the fluid-to-solid transition, both order parameters become long-ranged, the transition state (the critical nucleus)
has a more developed bond-orientational order, while the density is still comparatively low compared to the bulk crystal density (Fig.~\ref{fig:radial}).
Confocal experiments have the great potential of following the early stages of the nucleation process, and they do confirm a more rapid development
of orientational order over densification~\cite{tan2014visualizing,lu2015experimental,mohanty2015multiple}.

\subsubsection{Role of precursors in polymorph selection}

A natural extension of the ideas presented in the previous section is their application to
polymorph selection, which has attracted a lot of investigations~\textcolor{blue}{\cite{desgranges2006molecular,desgranges2007controlling,desgranges2006insights,kelton2010nucleation,zhou2011kinetics,russo_hs,bolhuis,russo_gcm,mithen2015nucleation,zhou2015crystallization,sosso2016crystal}}.

The observation that structuring foreshadows densification and crystallization leads to
the idea that these pre-structured regions (precursors) could play a role in
determining which polymorph is being nucleated. This idea requires the precursors to
have some underlying symmetry that is common with the crystalline structures that
are nucleated from them. Such behavior was observed for hard spheres, as reported in Ref. \cite{kawasaki,kawasaki2010correction,kawasaki2010structural}.

The first studies of polymorph selection from precursor regions were done with the Gaussian Core Model (GCM),
which is a good model for the effective interaction between the centers of mass of polymers dispersed in a good solvent~\cite{stillinger1976phase,likos2006soft}. The model presents two stable crystalline phases: the fcc
at low pressures, and the bcc phase at high pressures. The density of the bcc phase is lower
than the fcc phase, and it becomes lower than the density of the fluid phase above a certain pressure, after which
re-entrant melting is observed.
Nucleation in the GCM model was first considered in Ref.~\cite{bolhuis}, where the pre-structured cloud surrounding the nuclei
was taken into account to obtain a better description of the transition.
Ref.~\cite{russo_gcm} then proposed a link between the precursor regions and the crystal phase which was nucleated from them.
Contrary to thermodynamic predictions, it noted that the GCM model has a kinetic preference for the bcc phase, even in regions where the stable phase is fcc. Ref.~\cite{mithen2015nucleation}
then found that the nuclei have a mixed nature, not consisting of a single polymorph, and that the kinetic pathway selected during nucleation
persists even when the nucleus is many times above its critical size.

Another example comes from hard spheres. Here the relevant crystalline
structures are the close-packed fcc and hcp crystals. Despite a negligible bulk
free energy difference (of the order of $0.1\%$ of the thermal energy 
in favor of fcc~\cite{bolhuis_entropy,pronk}), simulations~\cite{luchnikov,snook,filion,russo_hs}
and experiments~\cite{gasser,pusey,palberg,jade_royall} have both found that nucleation predominantly
forms fcc. This preference for the fcc crystal form is indeed found in the precursor regions~\cite{russo_hs},
which show a clear preference for fcc-like local environments over the hcp-like ones.
This somehow highlights the idea that the kinetic pathway by which nuclei appear plays a fundamental
role in polymorph selection. The melt is not simply a structure-less background where nucleation can
occur homogeneously, but it is instead composed of local structures with different symmetries.
Symmetries that are compatible with long-range crystalline order serve as precursors, and their relative
abundance dictates the likelihood of formation of a specific polymorph.

Structures found in the supercooled melt are most often incompatible with crystalline order.
The most notable of these structures is the icosahedral-packing which,
since the pioneering work of Frank~\cite{frank1952supercooling}, is the archetypal model of amorphous structures.
While precursor regions are linked to nucleation, five-fold symmetric structures are linked to its avoidance.
In Fig.~\ref{fig:curves} we showed that, for hard-spheres at high density, the population of solid particles is outgrown by particles with icosahedral order~\cite{russo_hs}. Even a small fraction of icosahedral particles strongly suppresses the crystallization process~\cite{tanaka2003roles,karayiannis2011fivefold,laso_soft}. 
The effect is enhanced by polydispersity in particle sizes: for entropic reasons, polydispersity increases the
amount of icosahedral structures, where often the central particle is smaller than its neighbors.
At the same time, size asymmetry suppresses crystalline structures ~\cite{mathieu_russo_tanaka}.
Frustration effects of local structures of five-fold symmetry on crystallization were also discussed in two-dimensional single-component spin liquids \cite{ShintaniNP}.

The effects of polydispersity described here apply to close-packed crystalline structures. It was found that
open crystalline structures (as the structure of Ice in water) are instead comparatively more stable to disorder.
Angular disorder, in particular, was shown to have little effect on the crystallizability of systems with
directional interactions~\cite{romano2014influence}.

\subsubsection{Effects of external fields on precursor formation}
The effects of external fields on precursor regions were also considered~\cite{watanabe_walls,russo2013interplay,lander2013crystallization}.
External walls were considered experimentally and numerically in vertically vibrated quasi-two-dimensional granular liquids and two- and three-dimensional colloidal systems respectively~\cite{watanabe_walls}, showing that
the walls enhance structural order in the fluid, in a manner that is consistent with the bulk behavior, under an influence of wall-induced short-range translational ordering (i.e., layering).
This not only leads to the enhancement of glassy slow dynamics but also may induce heterogeneous nucleation near walls.
Ref.~\cite{russo2013interplay} instead considered the effect of rough walls on colloidal suspensions under the effects
of gravity, where the suppression of bond orientational order and layering next to the walls results in a strong suppression
of nucleation. 
Also shear was shown to suppress precursor regions~\cite{lander2013crystallization}, while at the same time increasing the rate
of growth of crystals. The interplay between these two effects is at the origin of the non-monotonous crystallization rate as a
function of the shear rate~\cite{lander2013crystallization}.

\subsubsection{Roles of precursors in vitrification}
Up to now we have addressed the role of structural heterogeneity in the melt. This not only promotes crystal nucleation, but also 
causes heterogeneous dynamics, which may be related to the so-called ``dynamic heterogeneities'', widely observed in glass-forming liquids. 
It is natural to expect that more structurally ordered regions are more stable and have slower dynamics than less ordered regions. 
The lifetime of these heterogeneous regions, $\tau_\text{het}$,
is comparable to/much longer than structural relation times for three-/two-dimensional systems respectively \cite{tanaka2012bond,tanaka2013importance}.
For hard-sphere systems (both mono and poly-disperse systems),
the study of dynamic heterogeneities in supercooled melts has showed that there is a close match between
slow regions and regions with high bond-orientational order~\cite{kawasaki,kawasaki2010correction,ShintaniNP,kawasaki2007correlation,tanaka,kawasaki2010structural,leocmach2012roles,mathieu_russo_tanaka,kawasaki2014structural,russo2015assessing,golde2016correlation}. 
This is suggestive of a possible general link between the glass transition and the growing length-scale of structural order, such as bond orientational order, at least for (nearly) single-component systems. A link between static and dynamic properties was found to apply also to system of monodisperse ellipsoids, where lenghtscales corresponding to glassy dynamics, heterogeneous dynamics, and structural order, where all found to diverge
at a common volume fraction for both translational and orientational degrees of freedom~\cite{zheng2014structural}.
We mention that in polydisperse emulsions a link between dynamic heterogeneities and local order was found, but the order corresponds to the
development of five-fold symmetric structures~\cite{zhang2016dynamical}. Further work will be needed to rationalize all these results
in a single framework.

Once a glass is formed, devitrification is still possible, but the mechanism is still not fully understood (see Sec.~\ref{sec:extending}),
despite the importance that nucleation from the glassy state has in many systems that have industrial applications, like pharmaceuticals and metallic glasses.
Recent investigations~\cite{sanz2011crystallization,sanz2014avalanches,taiki} suggest that the devitrification phenomena is linked to an intermittent
dynamic in the melt. During a series of discontinuous events, called \emph{avalanches}, a small fraction of particles undergoes rapid motion, which is
followed by an increase of the crystalline order in the melt.

A topic of practical importance in many applications is related to how the freezing transition is avoided~\cite{palberg2016make}, and what are the factors that determine the glass-forming ability
of a material. In Fig.~\ref{fig:curves} we saw that, at high density, the formation of crystalline particles is preempted by other structures that have a high local density.
These structures were identified as having five-fold symmetric symmetry, like icosahedral clusters~\cite{russo_hs}. It was also shown~\cite{mathieu_russo_tanaka} that, by
increasing polydispersity, these clusters are comparatively more stable than crystalline environments. It is thus clear that five-fold symmetric structures play an important
role in the avoidance of the crystallization transition \cite{frank1952supercooling,TanakaGJPCM,tanaka1999two,tanaka1999two2,tanaka2003roles}.
See Ref.~\cite{tanaka2012bond,tanaka2013importance} for an in-depth review of this topic. It further suggests an intrinsic link between crystallization and glass transition:
for example, 
suppressing the formation of crystal precursors may lead to higher activation energy for crystal nucleation and thus to higher glass-forming ability.
This topic is beyond the scope of this Article and will be discussed elsewhere~\cite{russo_eutectic}.

\subsection{Crystallization of liquids interacting with directional potentials: the case of water}
In the above, we consider precursors formed in systems interacting with isotropic potentials. 
In this case, local breakdown of rotational symmetry in a liquid state is only weakly coupled to local density, which,
in the initial stage of crystal nucleation, results in a stronger increase of
orientational ordering over the translational one. However, for a system interaction with a directional potential such as 
a tetrahedral liquid, the situation is very different since the symmetry selection by directional bonding 
automatically accompanies a local density change. The typical example can be seen in tetrahedral liquids such as water, Si, and Ge.
\begin{figure}
 \centering
 \includegraphics[width=8cm]{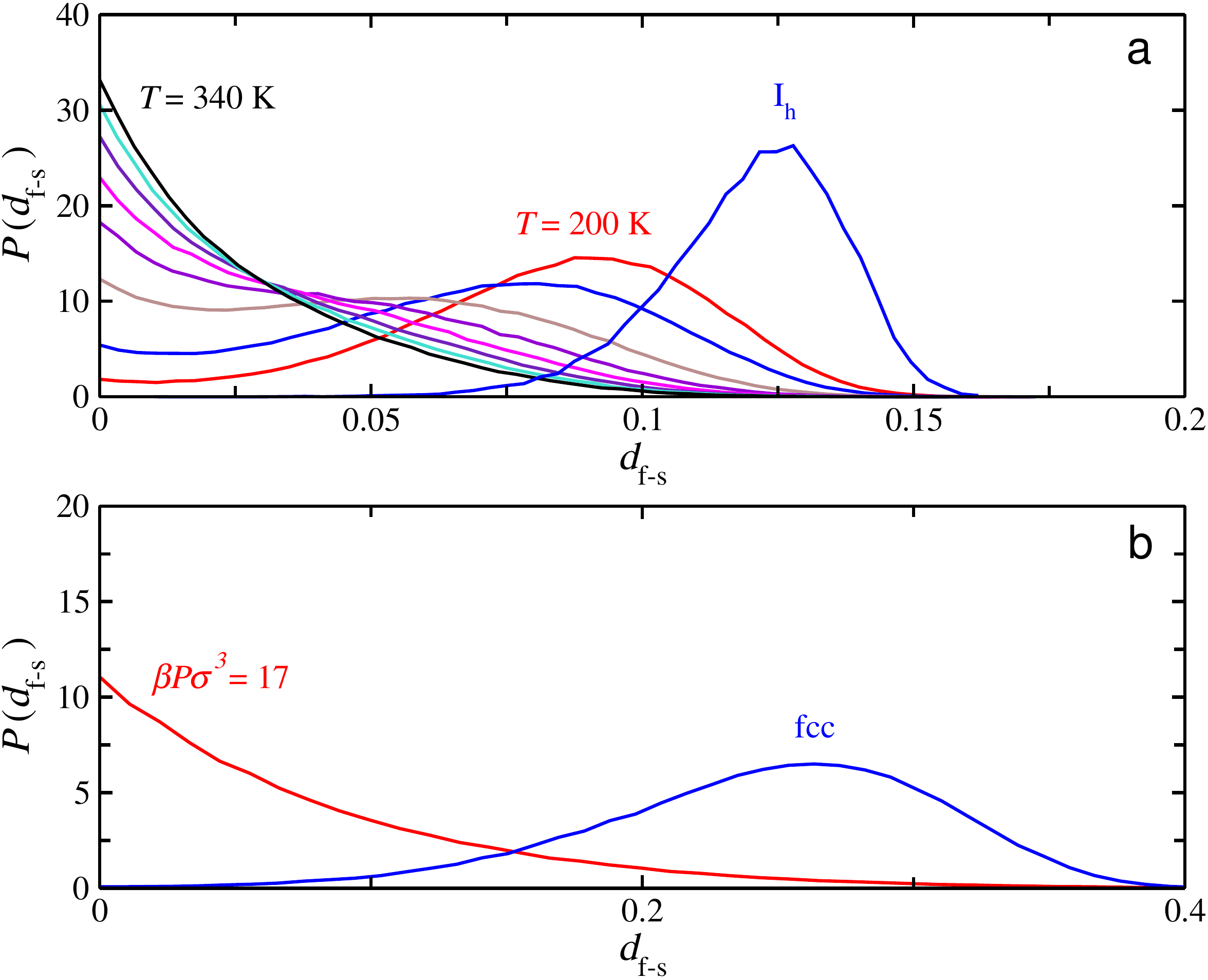}
  \caption{Positional order in water and hard spheres. Both panels show the probability distribution of the distance
 between the first and second shell, $d_{f-s}$, defined as the radial distance between the $(n+1)$-th and $n$-th nearest neighbour to the central atom.
 For water (panel a) we consider the oxygen atoms and $n=4$, while for hard spheres (panel b) $n=12$. (a) TIP4P/2005 water model at ambient pressure. Different curves correspond
 to temperatures ranging from $T=340\,K$ down to $T=200\,K$, with steps of $\Delta T=20\,K$. The blue $I_h$ curve refers to hexagonal $I_h$ at $T=200\,K$ and ambient pressure.
 (b) Hard sphere model, comparing the probability distribution of $d_{f-s}$ between the supercooled melt at $\beta P\sigma^3=17$ (red curve) and the fcc solid (blue curve). 
Reproduced from Ref.~\cite{russo2014understanding}.
 }
 \label{fig:firstsecondshell}
\end{figure}
In Fig.~\ref{fig:firstsecondshell} we compare the development of translational order in a popular molecular model of water,
TIP4P/2005~\cite{abascal2005general}, with the hard sphere system. Locally, translational order is a measure of the radial order
between pairs of atoms. In Fig.~\ref{fig:firstsecondshell} we plot the probability distribution of the distance between the
first and second shell of nearest neighbours, $P(d_{f-s})$, which is a measure of local translational order. In the case of water (Fig.~\ref{fig:firstsecondshell}(a)),
the probability distribution of $d_{f-s}$ shows an exponential decay at high temperatures, which is the expected distribution in a disordered environment. But as the
system is supercooled, the distribution function shows the development of states with a finite value of $d_{f-s}$, which signals the formation of the second shell.
At $T=200\,K$ the second shell is almost fully developed, and $P(d_{f-s})$ is close to the one measured in the ordered phase (the hexagonal $I_h$ ice).
Hard spheres, on the other hand, always have an exponential decay of $P(d_{f-s})$, up to the limit of homogeneous nucleation, Fig.~\ref{fig:firstsecondshell}(b), so that there
is little development of translational order prior to crystallization.

The different crystallization pathway, which involves the formation of locally favoured structures with a high degree of translational order,
is at the origin of thermodynamic and dynamic anomalies seen in such water-type liquids \cite{tanaka2012bond,tanaka2013importance,TanakaWPRB}, as will be shown in the next section.

\subsubsection{Local structural ordering and thermodynamic anomalies of water}

Water exhibits a surprising array of unusual properties that
mark it as ``the most studied and least understood of all known liquids''~\cite{stillinger_water}. 
Departures from the behaviour of simple liquids are often described as water anomalies, 
the most famous being the density maximum at about $4^\circ$C.
Most of these unique properties derive from hydrogen bonding, which is a strong directional interaction favoring
local tetrahedral arrangement of water in both its liquid and crystalline phases~\cite{Angell1983,Debenedetti2001,Debenedetti2003,moore2011structural,holten2013nature}.

Hints about the importance of structural ordering were first detected in the supercooled liquid region.
Here, the liquid-liquid
critical point (LLCP) scenario hypothesizes the existence of a second critical point in the metastable supercooled phase~\cite{poole1992phase,liu2009low,gallo_ising,poole_st2,palmer2014metastable,smallenburg2015tuning},
below which liquid water would phase separate into two different liquids, low-density liquid (LDL) and high-density liquid (HDL)  (the scenario has been intensely debated over the last years~\cite{limmer_chandler}, but recent results
seem to have confirmed it~\cite{palmer2014metastable,smallenburg2015tuning}).
Also glassy water has been found to exist in different states, called low-density (LDA), high-density (HDA) and very high-density (VHDA)~\cite{loerting2011many}
amorphous ices, which can interconvert with each other by the application of pressure. 
But also at ordinary conditions, many physical quantities exhibit behaviours suggestive of the presence of different states, 
such as infrared and Raman spectra \cite{Eisenberg,Walrafen_T}, X-ray absorption
spectroscopy, X-ray emission spectroscopy and X-ray small angle scattering~\cite{huang2009inhomogeneous,nilsson2011perspective}
(but these results are still debated~\cite{soper_gordon,saykally_gordon}),
and simulations~\cite{poole_mixture,wikfeldt_bimodal}. 
These phenomena directly point to the ability of the water molecule to sustain different local environments~\cite{gallo2016water,nilsson2015structural}.

This features have attracted considerable attention since the proposal of a mixture model by R\"ontgen \cite{Rontgen} and others 
(see, e.g., Refs. \cite{davis1965two,Angell_w,vedamuthu1994properties,ponyatovsky1998metastable,robinson1999isosbestic}). 
These models regarded water as a mixture of two rather distinct structures, whose entropy difference is not so 
significant.  
On a similar basis, but considering the formation of unique locally favored tetrahedral structures of low entropy 
in more disordered normal-liquid structures of high entropy, a two-order-paremter model was proposed \cite{tanaka2012bond,tanaka2013importance,tanaka1998simple,Tanaka2000a,tanaka2000simple}. 
In this model, the state of water is characterized by two order parameters, density and the number density of locally favored structures, $s$. In this framework, the liquid-liquid transition can be viewed as a gas-liquid-like transition of the order parameter $s$ 
\cite{tanaka2012bond,tanaka2013importance,TanakaLJPCM,tanaka2000general}. 
It was shown that modeling water as a mixture of two states has proved extremely successful in describing the anomalies of water
using a restricted number of fitting parameters~\cite{tanaka2000simple,holten_anisimov}, but their basic assumption, i.e.,
the existence of structurally different components still lacked a microscopic justification.
Here it may be worth pointing out that it may not be appropriate to use the terms LDL and HDL to express fluctuations in water, although this is popular.  For example, if we consider a system near its gas-liquid critical point, 
we characterize fluctuations in terms of the order parameter, density, and we do not refer low and high density regions to gas-like and liquid-like regions respectively. 

Now we re-examine water's unique properties from a microscopic viewpoint by focusing on its nucleation pathway, which we saw involved strong development of local
translational order prior to crystallization~\cite{russo2014understanding,russo2014new}. We focus here on a popular molecular model
for water, TIP4P/2005, which was especially designed to reproduce water's thermodynamic anomalies~\cite{abascal2005general,abascal2010widom}.

In the case of water, and tetrahedral materials in general, orientational order is best captured by an order
parameter called $Q_{12}$ which was introduced in Ref.~\cite{russo2014new} and shown to match the orientational
symmetry of all relevant crystalline structures with local tetrahedral coordination.

A good translational order parameter for water should then take into account the structure of
water up to the second nearest-neighbors shell, defined in terms of the network of hydrogen bonds. 
Common choices include the local structure index (LSI)~\cite{shiratani1998molecular,appignanesi},
the tetrahedral order parameter~\cite{Errington,limmer_molinero}, $g_5(r)$ (the average density of fifth-nearest neighbor)~\cite{poole_mixture}.
Recently we introduced a new structural parameter, $\zeta$ parameter~\cite{russo2014understanding}: for each water molecule, it is defined as the difference between
the radial distance of the closest oxygen in the second shell that is hydrogen bonded to the central water molecule, with the radial distance of the farthest oxygen atom in the first shell of neighbors. 
This $\zeta$ parameter includes information on
hydrogen bonding network in addition to the distance measure (see Ref.~\cite{russo2014understanding} on the details).
The probability distribution of the $\zeta$ parameter shows two distinct populations at supercooled conditions~\cite{russo2014understanding}.
The first population, with a high value of $\zeta$, is characterized by an open tetrahedral structure, while
the second population, with a small value of $\zeta$, is 
characterized by a collapsed second nearest-neighbor shell, with substantial shell interpenetration. Supercooled water structures can
thus be divided in two different states that differ for the structure of their second nearest-neighbor shell~\cite{soper2000structures,mishima_stanley}.
Locally favored states are represented by a Gaussian population centered around a finite positive value of $\zeta$, while the disordered state by
a Gaussian population centered around a null value of $\zeta$, with abundant number of states with shell interpenetration ($\zeta<0$). 
This bimodal distribution of the $\zeta$ parameter is in agreement with a two-order-parameter model of water \cite{tanaka2012bond,tanaka2013importance,tanaka1998simple,Tanaka2000a,tanaka2000simple,TanakaLJPCM,tanaka2000general}, in which 
water consists of two types of structures, normal liquid structures and locally favored structures and the fraction of the latter is the key order parameter $s$ 
specifying the state of water.

\begin{figure}[!t]
 \centering
 \includegraphics[width=8cm]{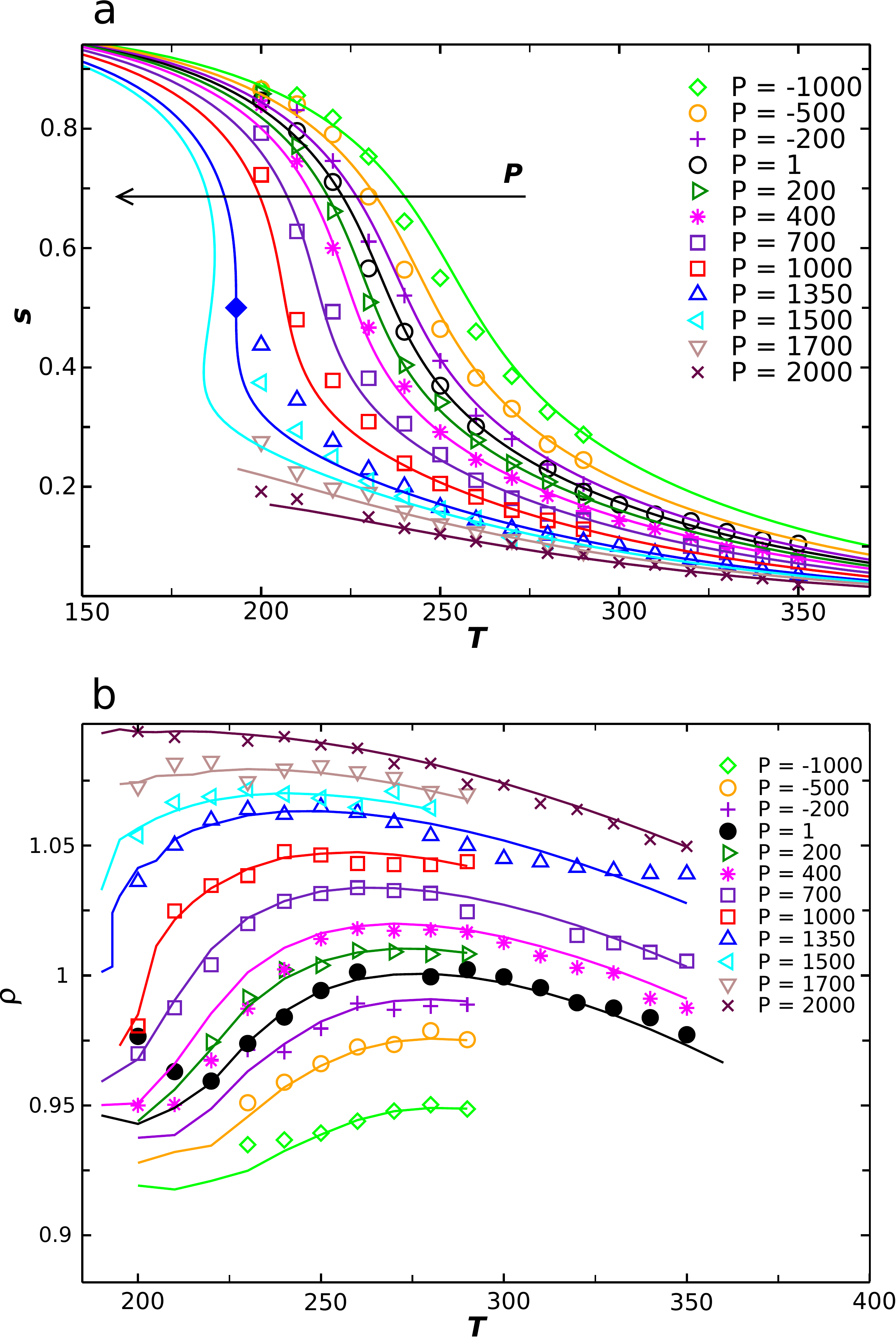}
  \caption{Two state model for TIP4P/2005 water. (a) Values of the fraction of the $S$ state ($s$) as a function of temperature for all simulated pressures. The symbols are the values obtained by the decomposition of the order parameter distribution, $P(\zeta)$, at the corresponding state point. Continuous lines are fits according to the two-state model.
 (b) Temperature dependence of density for several pressures. Continuous lines are simulation results, while symbols are obtained from the two-state model. Reproduced from Ref.~\cite{russo2014understanding}.}
 \label{fig:fig6}
\end{figure}

The advantage of the $\zeta$ parameter is that it allows a quantitative estimation of the fraction of low-energy (high $\zeta$) local structures directly
from computer simulation trajectories.
This fraction is reported as $s$ in Fig.~\ref{fig:fig6}(a) (symbols). 
The values of $s$ can be used for a two-state modeling of supercooled water, in which the free energy is written as a regular
mixture of two states, the high $\zeta$ and low $\zeta$ states~\cite{TanakaLJPCM,Tanaka2000a,russo2014understanding}. The prediction of the two state model is plotted with continuous
lines in Fig.~\ref{fig:fig6}(a). Once the free energy of the model is obtained in terms of $s$, it is possible to derive from it
all thermodynamic anomalies of water \cite{tanaka2012bond} 
and compare them with the ones measured in simulations.
For example, Figure~\ref{fig:fig6}(b) shows the comparison between simulations (lines) and two-state model predictions (symbols) for the density anomaly of a popular molecular model of water, TIP4P/2005 water \cite{russo2014understanding}.
The two-state model shows excellent agreement with the measured anomalies, proving that a microscopic two-state description of water's phase behavior
is possible. 
If there is a cooperative formation of such locally favored structure, the presence of two liquid phases, which are the gas and liquid state of locally favored structures, 
are expected at a certain condition~\cite{tanaka2012bond,poole1992phase,mishima_stanley,TanakaLJPCM,tanaka2000general,Tanaka2000a}, but whose direct observation is preempted by crystallization.

We have thus seen that structures with high values of $\zeta$ are directly responsible for
the anomalous behavior of water, and a description based on a two-state model is able to capture the intensity of
the anomalies in a quantitative way. The parameter $\zeta$ is an expression of local translational order, but structures
with high values of $\zeta$ are not crystalline structures. 
An analysis of the structures with high values of $\zeta$ in supercooled water reveals in fact that a large fraction of second nearest neighbors participates in five-membered rings of
hydrogen bonded molecules, and that this fraction increases with decreasing temperature and pressure~\cite{russo2014understanding}.
This five-membered rings are incompatible with the stable crystalline structures (both Ice Ih and Ic only have six-membered rings).
As the temperature decreases, the lifetime of hydrogen bonds increases,
and the opening of five-membered rings to form six-membered rings becomes increasingly more rare. This partly explains why water
has such a large metastability gap, in which, in absence of impurities, it can persist in its liquid form down to very low temperatures.

\subsubsection{Local structural ordering and homogeneous crystal nucleation in water}

The existence of the homogeneous crystallization line, where
the relaxation time of the liquid becomes shorter than the homogeneous crystallization time, 
implies that, at the best of today's knowledge, deeply supercooled liquid states will never be reached at bulk conditions, 
as already is found in simulations of coarse-grained water models~\cite{moore2011structural,limmer_chandler}.
Thus, the experimentally inaccessible supercooled liquid region is now widely known as ``no-man's land''~\cite{mishima_stanley}, 
although there have been efforts to overcome the difficulty by either strong spatial confinement~\cite{Mallamace,cerveny2016confined}, mixing an anti-freezing component~\cite{murata,murata2013general}, 
or rapid cooling \cite{Sellberg2014}. 

Homogeneous nucleation of ices has attracted considerable attention because of its importance in nature~\textcolor{blue}{\cite{sosso2016crystal}}. 
However, accessing it by numerical simulations has been very challenging particularly for realistic water models 
because of a large number of possible configurations of a disordered hydrogen-bond network, 
which makes the potential energy landscape complex.
The first homogeneous nucleation event in water at constant volume was reported for TIP4P \cite{matsumoto2002molecular}. 
It was found that ice nucleation occurs when a sufficient number of relatively long-lived hydrogen
bonds develop spontaneously at the same location to form a fairly compact initial nucleus.
At a high density, crystallization takes place more easily \cite{yamada2002interplay}. 
Ice nucleation was also reported by metadynamics simulations \cite{quigley2008metadynamics} 
and replica-exchange umbrella sampling \cite{brukhno2008challenges} in the isothermal-isobaric ensemble; 
however, the results may depend on the choice of order parameters.
Simulations using a coarse-grained mW water model \cite{molinero2008water} were also performed and shown to be 
very useful because of the relatively low computational cost \cite{li2011homogeneous,moore2011structural,moore2011cubic,reinhardt2012free}. 
For example, Moore and Molinero showed that a sharp increase in the fraction of four-coordinated molecules
in supercooled liquid water is responsible for water's anomalous thermodynamics
and the rapid increase in the rate of ice formation \cite{moore2011structural}. 
There are also many interests on how ice polymorphs are selected upon ice nucleation 
\cite{li2011homogeneous,moore2011cubic,malkin2012structure,zaragoza2015competition,haji2015direct}. 
Besides this problem of polymorph selection, the estimation of the barrier height and critical nucleus size 
is also an important issue and recently there have been many efforts toward this direction \cite{reinhardt2012free,sanz2013homogeneous,reinhardt2013note,espinosa2014homogeneous,espinosa2016seeding,haji2015direct,wang2015thermodynamics}. 
A link between the so-called Widom line \cite{Xu2005}, or the thermodynamic anomalies associated with the liquid-liquid transition,  
and the ice nucleation process was also suggested \cite{buhariwalla2015free}.

When we study crystal nucleation by numerical simulations, 
the choice of order parameters relevant for describing the miscroscopic pathway of crystal nucleation  
is crucial. In previous researches of ice nucleation, the bond orientational order parameter $Q_6$ 
and the hydrogen-bond network topology characterized by six-member rings are often used as key parameters. 
These choices are based on the assumption that the stable ice forms at ambient conditions are either hexagonal ice Ih or 
cubic ice Ic.  However, we recently showed \cite{russo2014new} that these order parameters may not be sufficient for describing 
the crystal nucleation pathway of water-type liquids \cite{TanakaWPRB}. Below we discuss this problem.
    
In order to crystallize, water structures need to acquire both translational and orientational order. 
Here we will briefly account for a recent scenario in which crystallization of ice near the homogeneous nucleation line, and in absence
of impurities, is influenced by a new metastable crystalline phase. The difference in the locally favored structure of supercooled
liquid water and the stable ice (Ice I) forms is a source of frustration of the crystallization process.
To overcome frustration effects, the pathway to crystallization can occur through intermediate steps, in line with Ostwald's step rule
of phases. In fact, in many molecular and soft-matter systems, crystallization does not occur directly in the stable crystalline phase,
but instead involves one or more intermediate steps where the melt crystallizes first in metastable phases.
These metastable crystals are structurally more similar to the melt than the stable phase. This structural similarity 
leads to a significant reduction of the interfacial energy, although the bulk free energy of metastable states is only intermediate between the one of the melt and of the stable phase.

This idea was recently put forward in Ref.~\cite{russo2014new}, where
a novel metastable phase was identified, called Ice 0, with a tetragonal unit cell with 12 molecules, whose thermodynamic and structural properties are intermediate between the melt and the solid crystalline phase.
The thermodynamic stability of the new phase was originally demonstrated for mW and TIP4P/2005 models of water~\cite{russo2014new}, and additional density functional calculations
have confirmed that, among the known metastable ice phases, Ice 0 has the closest free energy to that of stable ice I~\cite{slater2014crystal,quigley2014communication}.

\begin{figure}
 \centering
 \includegraphics[width=8cm]{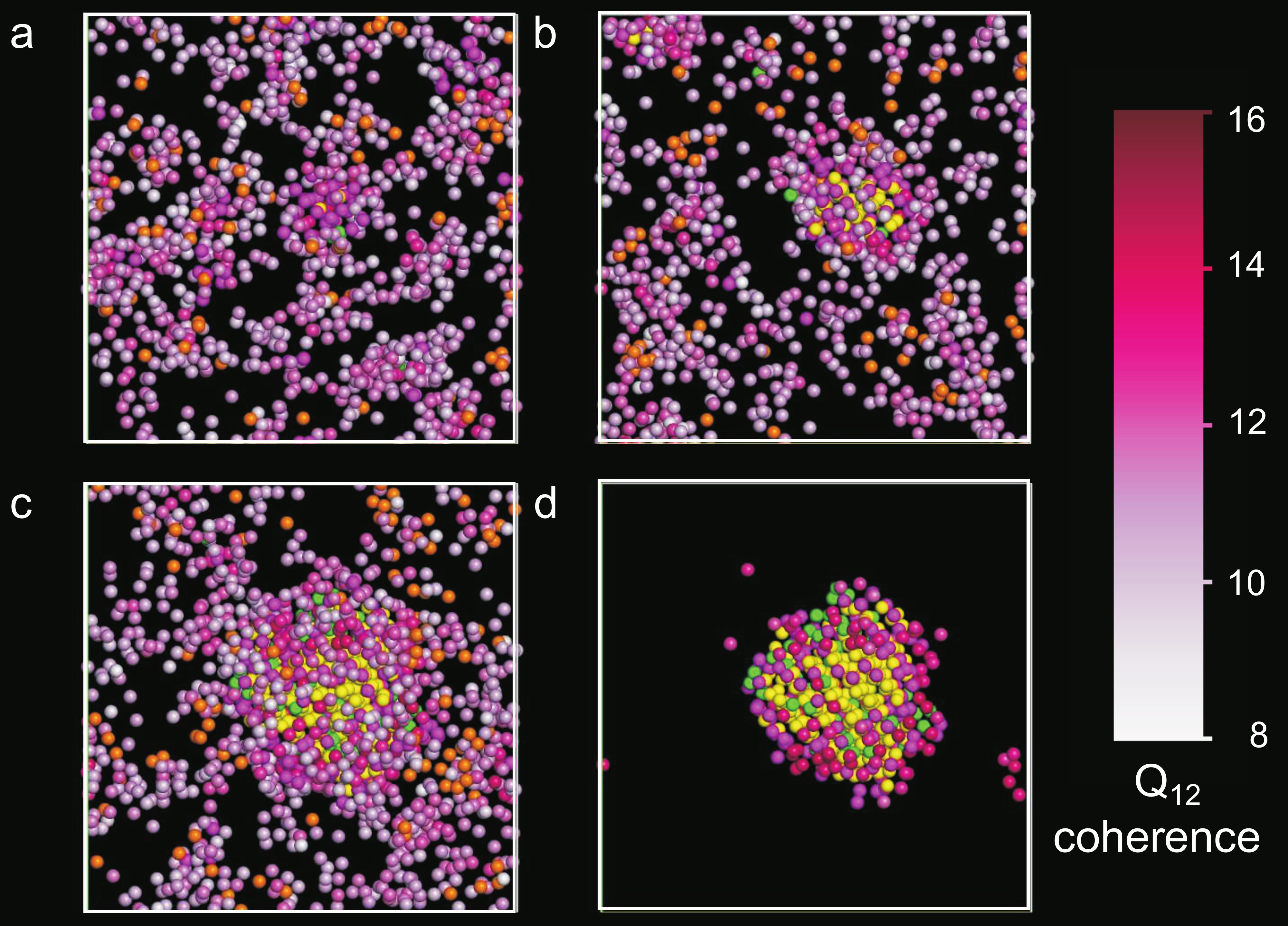}
 \caption{Snapshots from a crystallization trajectory at $T=206$ K and ambient pressure. The color code is the following: Ice Ic (yellow),
 Ice Ih (green), clathrate (orange), and Ice 0 (magenta). In (a)-(c) we also plot particles for which the number of connections is equal or bigger than 6.
 and color them according to their total coherence as shown in the color bar.
 (a) a pre-critical nuclei; (b) a nucleus of critical size ($n_c=80$); (c) the same nucleus at
 post-critical sizes. (d) the same as (c) but where only particles with 12 or more connections are plotted. 
 Reproduced from Ref.~\cite{russo2014new}.}
 \label{fig:fig7}
\end{figure}

A typical crystallization trajectory of mW water~\cite{molinero2008water} at $T=206$ K and ambient pressure is shown in Fig. \ref{fig:fig7} (see Ref. \cite{russo2014new} on the details).
The orientational order parameter $Q_{12}$ is used to detect precursor regions, where crystallization is likely to occur.
Precursors are colored in pink in the figure, and the color intensity increases with the amount of $Q_{12}$ order; the
stable Ic (cubic ice) and Ih (hexagonal ice) are colored in yellow and green respectively.
Precursor regions are not randomly distributed throughout the system, instead
they display a considerable amount of spatial correlation, which increases
with supercooling. Furthermore, Fig.~\ref{fig:fig7}(a) shows that the nucleation event happens inside the precursor regions.
This suggests that crystal nucleation is triggered by precursors in the supercooled liquid, which have a local symmetry 
consistent with that of the crystal to be formed. The idea that polymorph selection starts already in the supercooled liquid state is
essentially the same as the one discussed above for hard spheres \cite{russo_hs}.
In Fig.~\ref{fig:fig7}(b) we can see that an Ice Ic core has formed, and finally in Fig.~\ref{fig:fig7}(c), a post-critical nucleus is shown. 
In Fig.~\ref{fig:fig7}(d), where only crystalline particles (12 or more connections) are shown, 
we can see that indeed the surface is in contact with precursor regions.

\begin{figure}[!t]
 \centering
 \includegraphics[width=7.5cm]{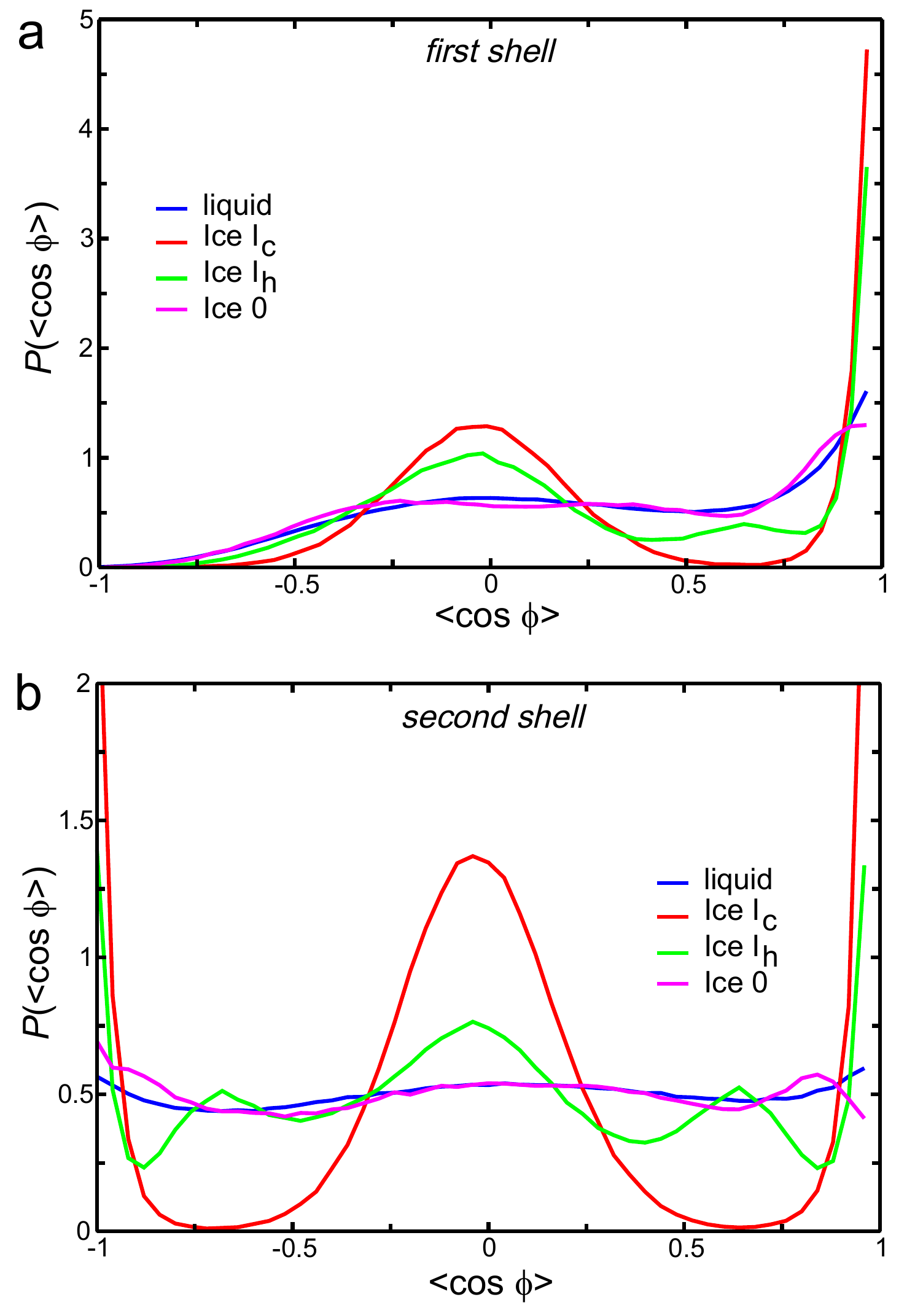}
 \caption{Distribution function of the average angle $\theta$ between the dipole moment of a water molecule
 and the dipole moment of its hydrogen-bonded nearest neighbours (panel a), and second nearest neighbours (panel b).
 Different curves correspond to TIP4P/2005 water, at $T=200$ K and $P=1$ bar, in the supercooled liquid (blue line),
 Ice $I_c$ (red line), Ice $I_h$ (green line), and Ice $0$ (magenta line) phases. See Supplementary Information of Ref.~\cite{russo2014new}
 for more details.  Reproduced from Ref.~\cite{russo2014new}.}
 \label{fig:dipoledipole}
\end{figure}

The idea to associate precursors with a metastable bulk phase (Ice 0) is based on the structural similarity between supercooled water and the Ice 0 phase. 
The similarity can be seen not only in local orientational symmetry, but also in the distribution of hydrogen-bonded ring structures: 
Both liquid water and Ice 0 have almost identical broad distributions of 5, 6, 7-membered rings peaked at 6, whereas Ice I is composed of 6-member rings only. 
The presence of 5 and 7 member rings acts as the source of frustration against direct crystallization into Ice I, but not for crystallization into Ice 0.
Another element of structural similarity is the dipole orientation between neighbouring water molecules, where supercooled liquid water and Ice 0 share
great similarities.
Fig.~\ref{fig:dipoledipole} plots the distribution function of the average angle $\theta$ between the dipole moment of a water molecule
 and the dipole moment of its hydrogen-bonded nearest neighbours (panel a), and second nearest neighbours (panel b). In both cases, it shows the remarkable
 similarity between the dipole-dipole local environment of supercooled liquid water and the Ice $0$ phase,while the stable phases Ice $I_h$ and $I_c$ show a very different
 local distribution of dipole moments.
As Ostwald's step rule of phases would suggest, nucleation could start in the intermediate Ice 0 form, and then converting to the stable crystalline
ices when nuclei are sufficiently big. 

\begin{figure}[!t]
 \centering
 \includegraphics[width=8cm]{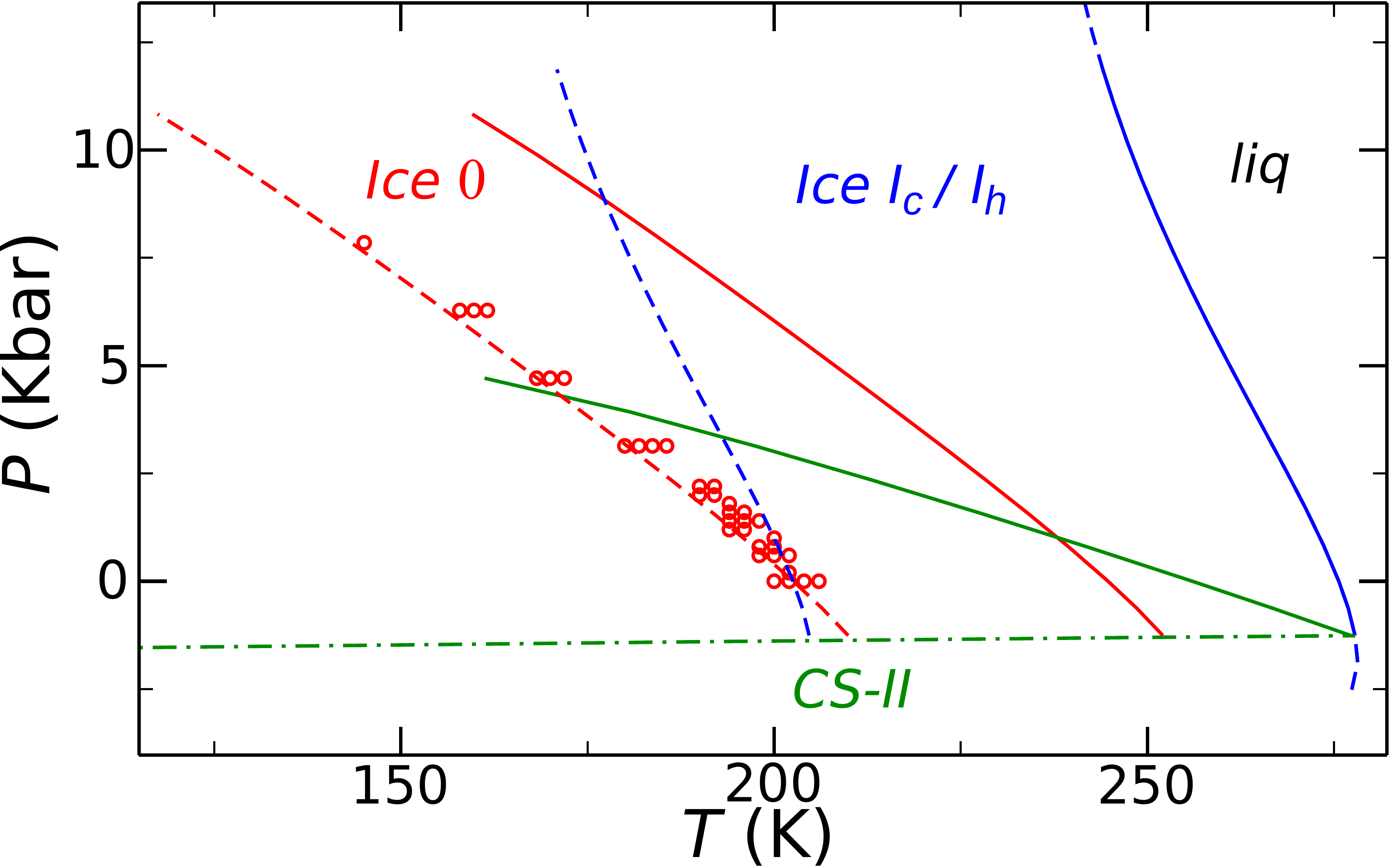}
  \caption{$P$-$T$ phase diagram of mW water. Continuous lines indicate coexistence between the liquid phase and different crystal structures: Ice Ih/Ic (blue), Ice 0 (red) and clathrate CS-II (green). Dashed lines indicate constant chemical potential differences between the liquid and Ice Ih/Ic ($\beta \Delta \mu = -0.721$, in blue), and the liquid and Ice 0 ($\beta \Delta \mu = -0.365$, in red). The green dashed-dotted line is the Ic/CS-II coexistence line. The red open circles indicate state points where homogeneous nucleation is observed in simulations.
 Adapted from Ref.~\cite{russo2014new}.}
 \label{fig:fig8}
\end{figure}

For example, in Fig.~\ref{fig:fig8} we report the phase diagram of the  mW model of water (a popular coarse-grained model of water).
Both melting lines of the stable Ice I phase (in blue) and of the metastable Ice 0 phase (in red) are depicted.
The homogeneous nucleation line, where the liquid phase looses its metastability, is represented with open (red) dots and it is seen coinciding with a line of constant thermodynamic driving force with respect to Ice 0
(i.e., the homogeneous nucleation line is the locus of constant chemical potential difference between the melt and Ice 0), rather than the stable phase Ice I.
In this scenario, it is shown that the presence of a metastable crystalline phase can fundamentally alter the pathway to nucleation in water.

\section{General importance of pre-ordering in our physical understanding of the liquid state}

In the above, we show that local structural ordering in a liquid state plays a significant role in both its static and dynamical behavior and non-equilibrium transitions taking place there, such as crystallization and vitrification. Conventional liquid-state theories are based on a physical picture that the liquid state is spatially homogeneous and disordered as in the gas state. 
This is of course true on a macroscopic scale, however, may not necessarily be the case on a microscopic scale, particularly at low temperatures. 
This local structural ordering cannot always be detected by the density order parameter and we need additional order parameters such as bond orientational order parameters. 
This indicates that the liquid state cannot be specified by the density field alone.  Such ordering must be driven by the free energy of the system, and structural 
ordering should be a consequence of these low free-energy structures. The many-body nature of this structural ordering is a direct consequence of rotational symmetry selection either by 
excluded volume effects under dense packing or by directional interactions that often have three-body nature.
As seen in the case of water, structural ordering can play a role also above the melting transition, the density anomaly at
$4^\circ$C being the most famous example of this. For ordinary fluids, dominated by repulsive interactions (such as the hard sphere model)
structural order seems to play an increasing role below the melting transition, and
a liquid-state theory would require the addition of these features to effectively describe the supercooled state. 
On noting that crystallization and vitrification take place in a supercooled state, such an extension 
may be essential for the description of these phenomena, as shown here (see also Ref. \cite{tanaka2012bond}).  
Here we note that the former-type ordering can be induced by excluded volume effects under dense packing for systems having rather hard repulsive interactions. 
For very soft systems, however, excluded volume effects may not be strong enough for symmetry selection due to large thermal symmetry fluctuations. 
So crystallization and vitrification phenomena may have different characters between soft and hard systems.

As reviewed above, pre-ordering in a liquid plays a crucial role in all kinds of phenomena taking place at rather low temperatures, such as 
crystallization, thermodynamic and dynamic anomalies of water-type liquids, liquid-liquid transition, and vitrification \cite{tanaka2012bond,tanaka2013importance}. 
Pre-ordered regions in a supercooled state act as precursors for crystal nucleation since the local symmetry of these regions should be similar to  
that of crystals because both symmetries are selected by the same free energy. This matching of the symmetry leads to a large reduction of the 
liquid/crystal interfacial energy. When there are more than two polymorphs, a polymorph whose symmetry is similar to precursors should be 
statistically favored upon nucleation. This should favor sequential ordering, known as the Ostwald step rule, in which such a metastable polymorph is first formed, and then transforms to another 
polymorph having a lower free energy.  However, it is worth noting that the selection rule explained above, which is based on the symmetry matching and the resulting  
low interfacial energy cost, is different from the original Ostwald step rule, which is based on the liquid/crystal free energy difference. 
Thermodynamic and dynamic anomalies of water-type liquids can be explained by 
a two-state model that assumes that the anomalies are caused by the formation of locally favored structures, which have static and dynamic properties 
different from normal liquid structures. 
Although we did not discuss liquid-liquid transition in this Article, it can be viewed as cooperative local structural ordering and the two liquid states can be regarded as 
the gas and liquid states of locally favored structures \cite{tanaka2012bond,tanaka2013importance}.
Finally, vitrification can be viewed as failure of crystallization because of frustration on 
crystalline preordering or competing ordering \cite{tanaka2012bond,tanaka2013importance}.

In this Article, we mainly focus on (nearly) single-component systems. 
For multi-component systems, which are a quite important class of materials, we should consider compositional order  
in addition to density and structural order. Even for binary mixtures, thus, we need at least three order parameters to describe their
liquid state. This is quite important to understand crystallization and vitrification in multi-component systems. 
This important and complex problem, in which one has to deal with a larger number of order parameters and with their couplings, is open for future investigation.

\section{Conclusions}
In the present Focus Article we examined nucleation as a multi-dimensional process in which the transition from
the liquid to the solid phase is driven by fluctuations in two (or more) order parameters. After a short introduction
to Classical Nucleation Theory, we showed that the first non-classical pathways to crystallization were discovered in the
context of two-step crystallization in models for proteins or colloids with short-range attractions. In two-step crystallization
scenarios, nucleation is aided by the the formation of metastable liquid droplets that considerably increase the crystallization rate.
This pathway was found to apply broadly for a variety of systems, from small molecules, to colloidal solutions, and protein solutions.
Even more surprisingly, the pathway of densification prior to nucleation was proposed to apply to systems outside the gas-liquid binodal, meaning that nucleation can occur even outside the region of metastability of the liquid droplets. This may be worth further investigation.

In the absence of a metastable liquid-gas phase separation, recent simulation and experimental work on colloidal systems have shown that, for spherically symmetric interactions, the precursor
field is not density but bond orientational order. Nucleation occurs first in regions with high bond orientational order, which then
compactify as crystallization proceeds. This is quite natural, on noting that an increase in orientational order is necessary for compactification of a dense system.
Moreover, with a high degree of orientational symmetry, the precursor regions can select the type of polymorph which will nucleate from it.
Several simulation results have shown that the polymorph that has been selected during nucleation can persist to sizes far bigger than the
critical size. The nucleation process,
i.e. the first stage of formation of tiny crystal nuclei, has a fundamental impact on the final state, or the macroscale structure.

We also touched upon the relation between precursor regions and heterogeneous dynamics in supercooled liquids, and the effect of external
fields (walls and shear) on these regions.

We then considered the case of water. Water has many unique thermodynamic and dynamic properties. The study of the nucleation pathway for both the translational and orientational field has shed new light on these properties. The translational order parameter field was described
in terms of a new structural parameter ($\zeta$) which was used to divide the population of water structures in two states.
A low-energy, high volume, ordered state (at high values of $\zeta$), and a high-energy, low volume and disordered state (at small values of $\zeta$).
The two-state model that derives from this microscopic description is able to quantitatively capture the anomalous behavior of water.
The orientational order of water is instead captured by another order parameter, $Q_{12}$. We have shown that the development of
orientational order, and the formation of small crystalline nuclei could be influenced by a new metastable phase, Ice 0, whose structure
is intermediate between that of the stable ice forms and of the supercooled liquid state. 

Finally, we discuss the general implications of this type of approach in our understanding of liquid-associated phenomena 
and mention important open problems. We hope that a liquid-state theory constructed on the basis of multiple order parameters 
and their static and dynamical couplings will be developed in the near future, and contribute to a better understanding 
of these fundamental open problems.

\section*{Acknowledgments}

This work was partially supported by Grants-in-Aid for Scientific Research (S) (21224011) and Specially Promoted Research (25000002) from the Japan Society for the Promotion of Science (JSPS). 
J.R. acknowledges support from the Royal Society through a University Research Fellowship.  

%\bibliography{biblio}


\end{document}